\documentclass[a4paper,11pt]{article}
\pdfoutput=1 

\usepackage{jheppub} 

\usepackage[T1]{fontenc} 

\usepackage{amsmath,amssymb}

\usepackage{braket}

\usepackage{bm}

\usepackage{here}

\usepackage{subcaption}

\usepackage{slashed}

\title{\boldmath Phase Transitions in Twin Higgs Models}

\author[a]{Kohei Fujikura,}
\author[b,c]{Kohei Kamada,}
\author[d]{Yuichiro Nakai}
\author[a]{and Masahide Yamaguchi}

\affiliation[a]{Department of Physics, Tokyo Institute of Technology,
2-12-1 Ookayama, Meguro-ku, Tokyo 152-8551, Japan}
\affiliation[b]{Center for Theoretical Physics of the Universe,
Institute for Basic Science (IBS), Daejeon 34126, Korea}
\affiliation[c]{Research Center for the Early Universe (RESCEU),
Graduate School of Science, The University of Tokyo, Tokyo 113-0033, Japan}
\affiliation[d]{Department of Physics and Astronomy, Rutgers University, Piscataway, NJ 08854, USA}

\emailAdd{fuji@th.phys.titech.ac.jp}
\emailAdd{kohei.kamada@resceu.s.u-tokyo.ac.jp}
\emailAdd{ynakai@physics.rutgers.edu}
\emailAdd{gucci@phys.titech.ac.jp}

\date{\today}

\abstract{
We study twin Higgs models at non-zero temperature and discuss
cosmological phase transitions as well as their implications on electroweak baryogenesis 
and gravitational waves.
It is shown that the
expectation value of the Higgs field at the critical temperature of the electroweak
phase transition is much
smaller than the critical temperature, which indicates two important
facts: (i) the electroweak phase transition cannot be analyzed
perturbatively (ii) the electroweak baryogenesis is hardly realized in
the typical realizations of twin Higgs models.  We also analyze the
phase transition associated with the global symmetry breaking, through
which the Standard Model Higgs is identified with one of the pseudo-Nambu-Goldstone bosons in terms of its linear realization, with and
without supersymmetry.  For this phase transition, we show that, only in
the supersymmetric case, there are still some parameter spaces, in which
the perturbative approach is validated and the phase transition is the first order. We find that the stochastic gravitational wave
background is generated through this first order phase transition, but
it is impossible to be detected by DECIGO or BBO in the linear realization and the decoupling limit. The detection of stochastic gravitational wave background with the feature
of first order phase transition, therefore, will give strong constraints
on twin Higgs models.}

\begin{document} 
\maketitle
\flushbottom

\section{Introduction}

Naturalness of electroweak symmetry breaking (EWSB) has been a
guideline for exploring physics beyond the Standard Model (SM).  Popular
scenarios of physics beyond the SM include supersymmetry (SUSY) and
composite Higgs, which are still promising solutions to the (large)
hierarchy problem, since they remove the sensitivity of the weak scale
to quadratically divergent quantum effects from physics at high energy
scales such as the Planck scale and the grand unification scale.
However, the discovery of the SM-like Higgs boson and nothing else at
the Large Hadron Collider (LHC) poses a problem for naturalness.
No new colored particles predicted in these popular scenarios have been
observed so far at the LHC, which already leads to fine-tuning in the
Higgs potential at sub-percent level. Although we do not know whether
nature takes thought for this little hierarchy problem or not, it is
interesting to pursue possibilities to ameliorate this fine-tuning
and to explore their implications for particle phenomenology and
cosmology.

The twin Higgs mechanism \cite{Chacko:2005pe} is an attractive idea to
provide a solution to the little hierarchy problem without introducing
new colored states.  There are several variations to realize this idea,
but every twin Higgs model starts with the assumption that the SM Higgs
field can be considered as one of the pseudo-Nambu-Goldstone bosons
(pNGBs) arising from spontaneous breaking of a global symmetry
${\cal G}$, such as $U(4)$ symmetry, that contains $SU(2)_A \times
SU(2)_B$ symmetry in its subgroups, to a smaller group ${\cal H}$, such
as $U(3)$.
Here $SU(2)_A$ and the mirror (or twin) $SU(2)_B$ are gauged and
interchanged under a (approximate) ${\bf Z}_2$ symmetry.  The
$SU(2)_A$ gauge symmetry is identified with the $SU(2)_W$ symmetry in
the SM and spontaneously broken by the vacuum expectation value (VEV)
of the Higgs field.  By introducing a $SU(3)_{\widehat{C}}$ mirror color
symmetry and mirror fermions that are charged under $SU(3)_{\widehat{C}}
\times SU(2)_B$, quadratic divergence to the Higgs potential coming from
the SM colored particles (and $SU(2)_W$ gauge bosons) are canceled by
the mirror colored particles (and $SU(2)_B$ gauge bosons).

The original realization of the twin Higgs idea, which is now called the
Mirror twin Higgs model \cite{Chacko:2005pe}, has a mirror copy of
all the SM particle content related to the ${\bf Z}_2$
symmetry.  On the other hand, the Fraternal (minimal) twin Higgs
model \cite{Craig:2015pha} has a smaller twin particle content,
that is, twin $W$ bosons, twin gluons and twin fermions corresponding to the third generation.
Other twin particles are not required since the corresponding SM particles give
less important contributions to the Higgs potential.
In any case, due to the ${\bf Z}_2$ symmetry, the quadratic terms of the Higgs potential accidentally preserve the original global symmetry ${\cal G}$
and the pNGBs associated with ${\cal G} \rightarrow {\cal H}$ breaking
are protected from radiative corrections, allowing the natural EWSB.
Since every twin partner is not charged under the SM gauge group,
this mechanism realizes the so-called neutral naturalness\footnote{
Another known realization of neutral naturalness is Folded SUSY \cite{Burdman:2006tz}.
}
and enables the model to evade stringent LHC bounds.
In this mechanism, there still remains the ``large hierarchy problem'',  
that is, the hierarchy between 
the global ${\cal G} \rightarrow {\cal H}$ breaking scale, expected to be up to 5-10 TeV, 
and more fundamental scales
such as the Planck scale or the grand unification scale. 
It is expected to be addressed by the UV completion
such as SUSY \cite{Falkowski:2006qq,Chang:2006ra,Craig:2013fga,Katz:2016wtw,Badziak:2017syq,Badziak:2017kjk,Badziak:2017wxn} or composite Higgs
\cite{Batra:2008jy,Geller:2014kta,Barbieri:2015lqa,Low:2015nqa,Csaki:2015gfd}. 

If the twin Higgs mechanism is really realized in nature, 
it may have significant impacts on cosmology since
the models predict new particles in the mirror (or twin) sector and have a rich structure in the Higgs sector.
One immediate concern in the Mirror twin Higgs model is the existence of mirror copies of light SM particles.
In fact, a twin photon and twin neutrinos give sizable contributions to the radiation energy density, which
is strongly disfavored by measurements of Cosmic Microwave Background (CMB) \cite{Ade:2015xua} and Big Bang Nucleosynthesis (BBN) \cite{Cyburt:2015mya}.
This issue has been recently studied in \cite{Barbieri:2016zxn,Chacko:2016hvu,Craig:2016lyx,Csaki:2017spo}.
The effects of twin baryons on the large scale structure and CMB 
are also investigated in Ref.~\cite{Chacko:2018vss}.
On the contrary, there is a candidate of dark matter in the Mirror twin Higgs model, which has been investigated 
in Ref.~\cite{Farina:2015uea}. 
The Fraternal twin Higgs model does not lead to an extra dark radiation component but still accommodates a dark matter candidate
\cite{Craig:2015xla,Garcia:2015loa,Freytsis:2016dgf,Prilepina:2016rlq}.

In addition to the modification in the relatively late-time
cosmology described above, we also naturally expect that the
thermal history of earlier Universe can differ from the standard one,
such as cosmological phase transitions, in the twin Higgs models.  For
example, in the SM without any extensions, the electroweak phase
transition is known to be crossover \cite{Csikor:1998eu}. Since the
Higgs sector is significantly different, it is non-trivial whether the
electroweak symmetry is really restored and how the EWSB proceeds even
if any. Kilic and Swaminathan addressed the first question and showed
that the electroweak symmetry is really restored at high temperature
\cite{Kilic:2015joa}.
In this paper, we try to address the second question, that is, how the EWSB
proceeds in the twin Higgs models.
If the electroweak phase transition is first order, which means that it proceeds through the bubble nucleation, then it is attractive in the cosmological point of view.
A first order phase transition generates stochastic gravitational wave (GW) background from bubble collisions~\cite{Turner:1990rc,Kosowsky:1991ua,Kosowsky:1992rz,Kosowsky:1992vn,Turner:1992tz}, sound waves~\cite{Hindmarsh:2013xza,Giblin:2014qia,Hindmarsh:2015qta,Hindmarsh:2017gnf}, and turbulence of the plasma~\cite{Kamionkowski:1993fg,Caprini:2006jb,Caprini:2009yp,Kosowsky:2001xp,Gogoberidze:2007an,Niksa:2018ofa}.
The typical peak frequencies of GWs originating from the first order phase transition associated with the EWSB are $\mathcal{O}(10^{-3}\sim 1)$ Hz, which are good targets of gravitational wave detectors such as DECIGO~\cite{Seto:2001qf} and BBO~\cite{Harry:2006fi}.
Therefore, if the twin Higgs models generally predict a first order phase transition associated with the EWSB, they can be tested by these observations.
Moreover, if the SM Higgs VEV at the critical temperature $T_C$, $\phi_A (T_C)$, is larger than the critical temperature, $\phi_A (T_C) /T_C >1$, inside the bubble, the electroweak phase transition is strong first order and the sphaleron decoupling condition is satisfied. (See App.~\ref{sec:thermal effective potential} for the definitions of a first order phase transition and a strong first order phase transition.) 
A strong first order electroweak phase transition accommodates electroweak baryogenesis~\cite{Trodden:1998ym,Morrissey:2012db}, so that the present baryon asymmetry of the Universe can be explained depending on the model parameters other than the (minimal standard) Higgs sector.

In this paper, we examine how the EWSB and the global symmetry breaking
${\cal G}\rightarrow {\cal H}$ (typically, $U(4) \rightarrow U(3)$)
proceed in thermal history of the Universe.  In particular, we address
the question if these phase transitions can be first order in
that framework.
We find that in the non-supersymmetric case, thermal
potential around both the electroweak and global symmetry breaking
cannot be analyzed perturbatively, which suggests that both phase
transitions are unlikely to be first order and hence we can
expect for neither the electroweak baryogenesis nor the generation of stochastic
gravitational wave background.  Even in the case with supersymmetric UV
completion, by limiting ourselves to the linear realization and the decoupling limit where only the
mirror stops are added to the non-supersymmetric model, we
find that the EWSB cannot still be analyzed perturbatively and the
conclusion is still robust. For the global symmetry breaking, however,
we show that, with an appropriate parameter choice, the symmetry
breaking can be first order and the associated
gravitational wave background is generated, but unfortunately it is too small to be detected by DECIGO or BBO. 

We organize this paper as follows. In Sec.~\ref{sec:twin}, we see
the electroweak vacuum structure of the non-supersymmetric twin
Higgs models as well as supersymmetric ones and discuss the fine-tuning
in the model parameters.
In Sec.~\ref{sec:twin phase}, we study twin Higgs
models with and without SUSY at non-zero temperature and examine how the EWSB proceeds.
In Sec.~\ref{sec:U(4) breaking phase transition}, we examine 
how the global symmetry breaking proceeds
and show that in supersymmetric twin Higgs models the first order
phase transition can be realized for appropriate parameter choices but the resultant gravitational wave background
is undetectable at planned gravitational wave detectors. 
Sec.~\ref{sec:discussion} is devoted to our concluding remarks and
comments.  In App.~\ref{app:finite temperature}, we exhibit the detailed
calculations and expression of the thermal potential and the spectrum of
the stochastic wave background from a first order phase
transition.

\section{Twin Higgs Models}\label{sec:twin}

In the first part of this section, we review the twin Higgs mechanism,
which provides a solution to the little hierarchy
problem~\cite{Chacko:2005pe,Craig:2015pha}.
The Higgs mass formulae are also presented.
We then discuss the degree
of fine-tuning to realize the adequate EWSB in this scenario.
In the second part, we describe a supersymmetric realization of the twin Higgs mechanism.

\subsection{The non-supersymmetric twin Higgs}\label{sec:minimal twin Higgs}

In the twin Higgs mechanism, the SM Higgs field is identified with pNGBs
arising from spontaneous breaking of an approximate $U(4)$ symmetry
(explicitly broken by the Yukawa and gauge
couplings).\footnote{Here, we confine a global group $G$ to $U(4)$
symmetry as a concrete realization. However, our conclusion is still
robust even for a generic gauge group $G$ as long as we consider a Mexican-hat type potential given by \eqref{twinhiggspotential}.} Let us consider a linear realization of the
mechanism and write a $U(4)$ symmetric potential of a complex scalar
field $H$ with the fundamental representation,
\begin{align}
V(H) = -m^2 H^\dagger H + \lambda \left( H^\dagger H \right)^2 ,\label{twinhiggspotential}
\end{align}
where $\lambda >0$ is required from the stability of the potential.
This potential drives the scalar field $H$ to get a nonzero VEV, $f
\equiv \langle |H| \rangle = m / \sqrt{2 \lambda}$.  Then, the global
$U(4)$ symmetry is broken down to $U(3)$ yielding $7$ NGBs.  The $U(4)$
symmetry contains the subgroups $SU(2)_A \times SU(2)_B$ and the scalar
field can be decomposed as $H = (H_A, H_B)$, where $H_A$ transforms
as a doublet of $SU(2)_A$ while $H_B$ does as a doublet of $SU(2)_B$.
$H_A$ is identified with the SM Higgs doublet and the $SU(2)_A$
symmetry is regarded as the ordinary $SU(2)_W$ gauge symmetry.  The
$SU(2)_B$ symmetry is gauged and becomes the twin $SU(2)_{\widehat{W}}$.
Then, the 6 pNGBs are eaten by the gauge bosons after the symmetry breakings while the remaining
one is the observed SM-like Higgs boson $h$.  A physical heavy exotic
Higgs $\widehat{h}$, corresponding to the radial direction, has the
mass $m_{\widehat{h}} = \sqrt{2 \lambda} f$ from Higgs mechanism.
As described in the introduction, $SU(2)_A$ and the twin $SU(2)_B$ are
interchanged under a ${\bf Z}_2$ symmetry.  In the Fraternal twin Higgs
model \cite{Craig:2015pha}, only the ${\bf Z}_2$ partners of the third generation of quarks and leptons and the partners of gluons (twin gluons) as well as the twin
$SU(2)_{\widehat{W}}$ gauge bosons are introduced.  Then, the twin Higgs
doublet $H_B$ has the following Yukawa coupling similar to the SM top
Yukawa coupling,
\begin{align}
\mathcal{L} \supset -\widehat{y}_t  H_B \widehat{Q}^a \widehat{t}_R^a + \rm h.c. \, ,
\end{align}
where $\widehat{Q}^a$ ($a =1,2,3$) are twin left-handed top (bottom)
quark doublet charged under the twin $SU(3)_{\widehat{C}}$ and
$\widehat{t}_R^a$ are twin right-handed top quarks.  $\widehat{y}_t$ is
the twin top Yukawa coupling whose value is almost the same as the
ordinary top Yukawa coupling $y_t$ due to the approximate ${\bf
Z}_2$ symmetry.

The two scalar doublets $H_A$ and $H_B$ receive quadratically divergent
corrections from the top and twin top quarks respectively as well as
corrections from the $SU(2)_A$ and $SU(2)_B$ gauge bosons at one loop.
In addition, they receive corrections from the gluons and twin gluons at
two-loop level. The quadratically divergent part of their potential is
given by
\begin{align}
V \supset \left( -\frac{3y_t^2}{8\pi^2}+\frac{9g_2^2}{64\pi^2} -\frac{3 y_t^2 g_3^2}{8\pi^4}  \right)
\Lambda^2 |H_A|^2
+\left( -\frac{3\widehat{y}_t^2}{8\pi^2}+\frac{9\widehat{g}_2^2}{64\pi^2}
-\frac{3 \widehat{y}_t^2 \widehat{g}_3^2}{8\pi^4} \right)\Lambda^2 |H_B|^2,
\label{eq:quadraticterm}
\end{align}
where $g_2, \, \widehat{g}_2$ are the $SU(2)_W$ and
$SU(2)_{\widehat{W}}$ gauge couplings, $g_3, \, \widehat{g}_3$ are the
$SU(3)_C$ and $SU(3)_{\widehat{C}}$ gauge couplings and $\Lambda$ is a
cutoff scale. The exact ${\bf Z}_2$ symmetry leads to
$\widehat{y}_t= y_t, \, \widehat{g}_2= g_2, \, \widehat{g}_3= g_3$ which
guarantee that the quadratically divergent part of the potential
respects the full $U(4)$ symmetry.  Then, the NG nature of the Higgs
field is not explicitly broken by the quadratically divergent
corrections, addressing the little hierarchy problem.
However, the SM Higgs would be exactly massless and inconsistent with our Universe
if the $U(4)$ and ${\bf Z}_2$  symmetries are exact. 
Thus we need small breakings of these symmetries.

Let us consider the breaking of the $U(4)$ and
the ${\bf Z}_2$ symmetries to give the appropriate effective Higgs potential.
First of all, the gauged $SU(2)_A \times SU(2)_B$ group has already broken the 
$U(4)$ symmetry explicitly.
In addition to the quadratically divergent
corrections, this generates logarithmically divergent contributions to the
quartic couplings of the form $\left( |H_A|^4 + |H_B|^4 \right)$, which
do not respect the $U(4)$ symmetry 
and then contribute to the Higgs
boson mass.  The explicit ${\bf Z}_2$ symmetry breaking is also needed
otherwise the hierarchy, $v_A^2 \ll f^2$,  
is not fulfilled, which is required
to satisfy the constraint from the Higgs coupling measurement.
We do not specify a mechanism to generate this breaking in this paper, but just encapsulate
the breaking effect in the quadratic and quartic terms of $H_A$.  The
effective potential of the scalar field $H$ we consider here is then summarized as
\begin{align}
V =\lambda\left(|H_A|^2+|H_B|^2-\frac{f^2}{2} \right)^2
+\kappa_1 \left(|H_A|^4+|H_B|^4 \right)+ \sigma_1 f^2 |H_A|^2 +\rho_1 |H_A|^4.  \label{eq:general}
\end{align}
The first term is the $U(4)$ conserving term coming from the original potential Eq.~\eqref{twinhiggspotential} rewritten in terms of
$H_A$ and $H_B$ and the corrections in Eq.~\eqref{eq:quadraticterm}, which determines the $U(4)$ symmetry breaking scale $f$. 
The second term that breaks the $U(4)$ symmetry includes the 
gauge (and top Yukawa) contributions in the Coleman-Weinberg potential. 
Thus $\kappa_1$ will be of order $g_2^4 /16 \pi^2 \log \left( \Lambda / g_2
f\right)$.
The third and
fourth terms are the ${\bf Z}_2$ breaking terms.  
The third term is induced, {\it e.g.,} by the quadratic corrections with ${\bf Z}_2$-breaking
part in the gauge and matter sector.
$\rho_1$ in the fourth term  includes the contribution of the one-loop
Coleman-Weinberg potential. In the Fraternal twin
Higgs model, the fourth term could arise from the ${\bf Z}_2$ breaking
effect such as the absence of the $U(1)_Y$ gauge symmetry in the twin sector.
However, this effect is of order $g_1^4 /16\pi^2$,
where $g_1$ is the $U(1)_Y$ gauge coupling constant and tiny.
In summary, we take $\lambda, f, \kappa_1, \sigma_1$ and $\rho_1$ to be the model parameters  
and require
$\sigma_1,~\kappa_1,~\rho_1 < \lambda$ so that the second, third and the forth terms in Eq.~\eqref{eq:general} are regarded as perturbations to the first term.

 At energies well below the symmetry breaking
scale $f$, we can integrate out the Higgs field $H_B$, which enables us to work with an
effective field theory of the SM Higgs field $H_A$.  The effective
potential of the SM Higgs field can be obtained by setting $H_B$ as
\begin{align}
|H_B|^2 =\frac{f^2}{2} -|H_A|^2. \label{eq:EFT}
\end{align}
Using this relation, we find
\begin{align}
V_{\rm eff} (H_A) = -(\kappa_1 - \sigma_1) f^2 |H_A|^2 +(2 \kappa_1 + \rho_1) |H_A|^4  . \label{eq:sm}
\end{align}
This potential coincides with the SM Higgs potential when the parameters
$\kappa_1$, $\sigma_1$ are identified with
\begin{align}
2 \kappa_1+\rho_1 = \lambda_{\rm SM}, \qquad  \frac{\kappa_1 - \sigma_1}{2\kappa_1 +\rho_1} = \frac{v_A^2}{f^2} , \label{eq:nonlinear}
\end{align}
where $\lambda_{SM} \sim \frac{1}{8}$ is the SM Higgs self-coupling,
$v_A = 246 \, \rm GeV$ is the VEV of the Standard Model Higgs
field. As denoted above, to satisfy the constraint from the Higgs coupling measurement,
the VEV of the Standard Model Higgs field is required to be satisfactorily small compared to the $U(4)$ symmetry breaking scale, that is, $v_A^2 \ll
f^2$.

Let us discuss the EWSB conditions ($v_A \simeq 246$GeV and $m_h \simeq 125$GeV) in the twin Higgs
models precisely with the potential~\eqref{eq:general}.  
By expressing the potential \eqref{eq:general} in terms of the two physical modes $\phi_A$ and $\phi_B$ with 
$H_A \equiv (0, \phi_A/\sqrt{2})$ and $H_B \equiv (0, \phi_B/\sqrt{2})$
and requiring the minimization
conditions, $\partial V/ \partial \phi_A =
\partial V / \partial \phi_B = 0$, we find the potential minimum given by 
\begin{align}
v_A^2 =  \lambda f^2 \frac{-\sigma_1 +\kappa_1 (1-\frac{\sigma_1}{\lambda})}{\lambda \rho_1 +\kappa_1 (2\lambda +\rho_1 +\kappa_1)} .  \label{eq:SM VEV}
\end{align}

Evaluating the mass matrix $\partial V/ \partial \phi_i \partial \phi_j (i,j=A,B)$ 
around the potential minimum,
we obtain the mass eigenvalues of the system, that is, the SM Higgs boson $h$ and the heavy
exotic (global symmetry breaking) 
Higgs $\widehat{h}$ as \cite{Katz:2016wtw,Badziak:2017syq}
\begin{equation}
\begin{split}
m^2_{\widehat{h},h}&=\rho_1 v_A^2 +f^2(\lambda+\kappa_1)\left(1\pm \sqrt{1-A}\right),\\[1ex]
A&\equiv 2\frac{v_A^2}{f^2} \frac{\lambda \rho_1 +\kappa_1(4\lambda+\rho_1+2\kappa_1)}{(\lambda+\kappa_1)^2}-\frac{v_A^4}{f^4}\frac{4\lambda \rho_1 +\rho_1^2 +\kappa_1 (8\lambda +4\rho_1 +4\kappa_1)}{(\lambda +\kappa_1)^2}, \label{exacthiggsmass}
\end{split}
\end{equation}
where the plus sign in front of $\sqrt{1-A}$ corresponds to
$m^2_{\widehat{h}}$ and the negative sign corresponds to $m_h^2$.  With
$v_A^2 / f^2 \ll1$, the SM Higgs mass is approximately given by
\begin{align}
m_{h}^2 \simeq 2 \frac{\kappa_1^2 + 2 \kappa_1 \lambda + \kappa_1 \rho_1 + \lambda \rho_1}{\kappa_1 + \lambda}
v_A^2 \, . \label{eq:SM-like Higgs mass}
\end{align}

\begin{figure}[t]
\centering\includegraphics[width=7cm]{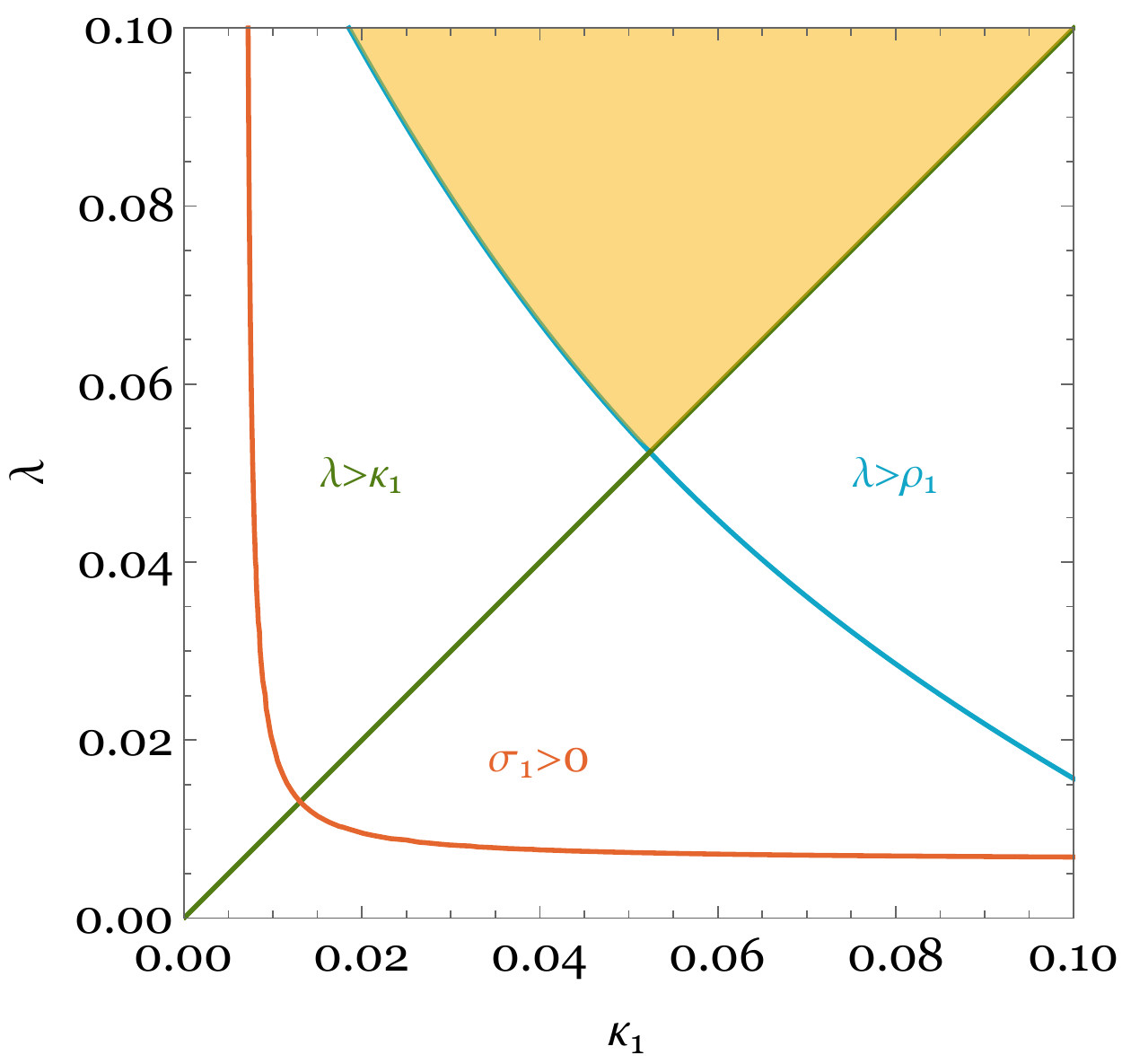}
\centering\includegraphics[width=7cm]{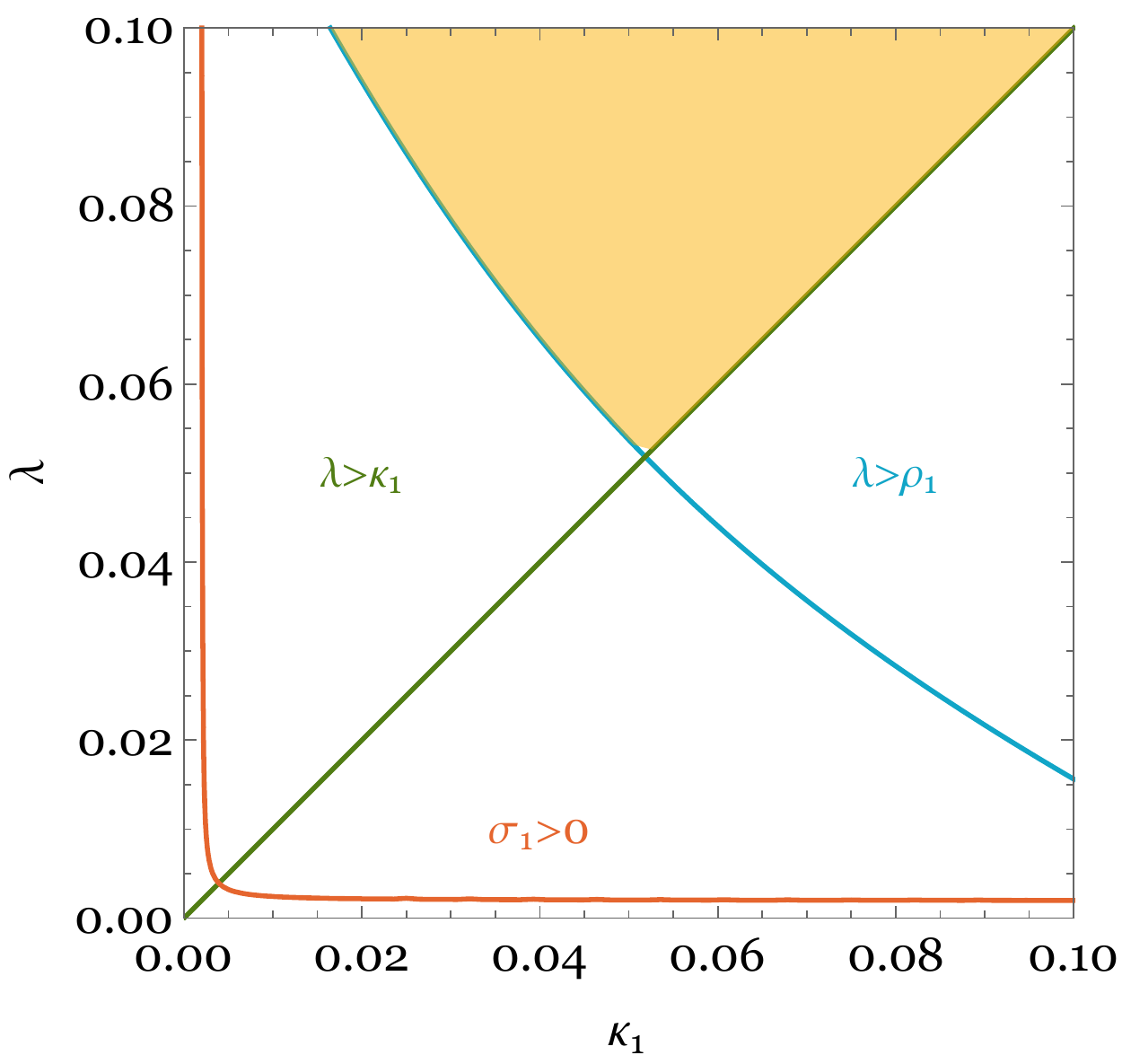}
\caption{The (yellow) region that satisfies 
$\lambda > \rho_1,~\kappa_1$ and $\sigma_1 >0$ is shown for $v_A \simeq 246$ GeV,  $m_h\simeq 125$GeV, 
and $v_A  / f = 0.223$ (left) and 0.123 (right).  
The regions above the blue, green and red curves satisfy with the constraints $\lambda > \rho_1,~\lambda>\kappa_1$ and $\sigma_1 >0$, respectively.
}\label{fig:allowed_region}
\end{figure}

Since we have five parameters $f,~\lambda,~\sigma_1,~\kappa_1$ and $\rho_1$,  
after imposing the EWSB conditions $v_A \simeq 246$ GeV \eqref{eq:SM VEV} and  $m_h \simeq 125$GeV \eqref{exacthiggsmass}, the system is now described by three parameters.
As noted above, we impose the conditions $\lambda > \sigma_1, \kappa_1, \rho_1$ to keep the philosophy of the twin Higgs 
models.
Fig.~\ref{fig:allowed_region} shows the parameter space that satisfies these conditions for $v_A / f = 0.223$ ($f=1.1$ TeV) and 0.123 ($f=2$ TeV). 
We also confirmed that the condition $\lambda > \sigma_1$ is always satisfied.
Note that the parameters $\lambda,~\kappa_1$ and $\rho_1$ cannot take arbitrary small values because tiny $\lambda,~\kappa_1$ and $\rho_1$ cannot realize the SM-like Higgs mass.
In fact, we can see from Fig.~\ref{fig:allowed_region} that the smallest values of $\lambda,~\kappa_1$ and $\rho_1$ are roughly given by $\lambda \simeq \kappa_1 \simeq \rho_1 \simeq 0.05 $.
This bound will play important roles when we analyze the dynamics of a phase transition as we will see in Sec.~\ref{sec:U(4) breaking phase transition}.
The smallest values of $\lambda,~\kappa_1,~\rho_1$ are not sensitive to the breaking scale $v_A / f$ and SM-like Higgs mass $m_h\simeq 125$GeV.
$\sigma_1 > 0 $ guarantees our assumption of the two-step phase transition as we will see later.

Let us finally examine the fine-tuning in this effective potential.
We estimate the degree of tuning by
the measure defined in Ref.~\cite{Barbieri:1987fn},
\begin{align}
\Delta (p_i) \equiv \left| \frac{\partial \log O(p_i)}{\partial \log p_i} \right|^{-1}, 
\end{align}
where $p_i$ are the model parameters and $O(p_i)$ are observables.
In this measure, smaller $\Delta (p_i)$ means that larger fine-tuning is required.
Thus $\Delta (p_i)$  should not be too small for the naturalness, say,  
at least all the measures should satisfy $\Delta (p_i) >{\cal O}(10^{-2})$. 
If a measure in the model is too small, $\Delta (p_i) \leq {\cal O}(10^{-2})$, 
we conclude this model is unnatural.
In our effective potential, the set of the observable and parameter that gives 
the smallest measure is the following one, 
\begin{align}
\Delta_{\sigma_1} &\equiv \left| \frac{\partial \log \left( v_A^2 / f^2 \right)}{\partial \log  \sigma_1}\right|^{-1}
= \frac{2\frac{v^2_A}{f^2}}{1-2\frac{v^2_A}{f^2}}\simeq \frac{2v_A^2}{f^2}.
\end{align}
In this calculation, we simply assume the soft breaking scenario, $\sigma_1 \gg \rho_1$, which means that the twin ${\bf Z_2}$ symmetry is only broken by the soft term $\sigma_1 f^2$.\footnote{For the hard breaking scenario, $\rho_1 \gg \sigma_1$, see Ref.~\cite{Katz:2016wtw}.}
In order to solve the little hierarchy problem in twin Higgs models,
$\Delta_{\sigma_1}$ should not take an arbitrary small value. Thus,
the symmetry breaking scale $f$ is bounded from above in light of
naturalness.

\subsection{Supersymmetric twin Higgs models}\label{sec:SUSY twin Higgs}

To address the large hierarchy problem in the twin Higgs scenario, SUSY
can provide an attractive solution. Parallel to the case of the
ordinary Minimal Supersymmetric Standard Model (MSSM), where Higgs
chiral multiplets consist of a pair of doublets, supersymmetric
twin Higgs models generally contain four Higgs doublets,
\begin{align}
H_u=
\begin{pmatrix}
H_u^{A}\\[0.5ex]
H_u^{B}
\end{pmatrix}, \qquad
H_d=
\begin{pmatrix}
H_d^{A}\\[0.5ex]
H_d^{B}
\end{pmatrix}.
\end{align}
The chiral multiplets $H_u$, $H_d$ are fundamental under the $U(4)$
symmetry and the $U(4)$ multiplets are decomposed into the visible
sector fields $H_u^A$, $H_d^A$ and the twin sector fields $H_u^B$,
$H_d^B$ under the subgroups $SU(2)_A \times SU(2)_B$.  The
superpotential contains an extended version of the ordinary $\mu$-term,
$W \supset \mu (H_u^A H_d^A + H_u^B H_d^B)$.  Including soft SUSY
breaking mass terms, the quadratic part of the $U(4)$ symmetric
potential in supersymmetric twin Higgs models is given by
\begin{equation}
\begin{split}
V_{U(4)} \supset \,& \left(\widetilde{m}^2_{H_u} +\mu^2 \right) \left(|H_u^A|^2 +|H_u^B|^2 \right) + \left(\widetilde{m}^2_{H_d}
+\mu^2 \right) \left(|H_d^A|^2+|H_d^B|^2 \right) \\
&-b \left(H_u^A H_d^A + H_u^B H_d^B +{\rm h.c.}  \right) .
\end{split}
\end{equation}
The quartic part of the $U(4)$ symmetric potential is model dependent.
The first term of \eqref{eq:general} contains the quartic term $|H_A|^2
|H_B|^2$.  However, SUSY forbids this type of couplings without further
modification of the Higgs sector.  There are several proposals to obtain
Higgs couplings with twin Higgs fields.
Refs.~\cite{Falkowski:2006qq,Chang:2006ra,Craig:2013fga,Katz:2016wtw}
have introduced a massive singlet chiral superfield $S$ with a
superpotential $SH_u H_d$. The effective theory after integrating out
this singlet contains the quartic term $|H_u H_d|^2$.
Ref.~\cite{Badziak:2017syq} has considered an additional contribution to
the $D$-term potential from a new $U(1)$ gauge symmetry, under which
both the Higgs and the twin Higgs fields are charged.  In this paper, we
do not go into the details of a specific supersymmetric twin Higgs
model, instead, we simply assume the existence of an appropriate $U(4)$ symmetric 
quartic term and try to extract general features of a
supersymmetric twin Higgs scenario.

We next consider possible sources of the breaking of the $U(4)$ and
the ${\bf Z}_2$ symmetries in this scenario.  In the non-supersymmetric
minimal model, the $U(4)$ symmetry breaking arises only from
quantum corrections or from some explicit breaking terms.  On the
other hand, supersymmetric models have the $D$-term potential,
\begin{align}
V_{D} = \frac{g_1^2+g_2^2}{8} \left(|H_u^A|^2-|H_d^A|^2 \right)^2 +\frac{\widehat{g}_2^2}{8} \left(|H_u^B|^2 -|H_d^B|^2 \right)^2 ,
\end{align}
which breaks the $U(4)$ and the ${\bf Z}_2$ symmetries.  Here we have
assumed the minimal realization of the twin Higgs mechanism, where the
twin partner of $U(1)_Y$ is not introduced.  Unfortunately, the size of the ${\bf Z}_2$ breaking in the
above $D$-term potential is insufficient to realize the required
hierarchy between the electroweak breaking scale, $v_A$, and the $U(4)$
breaking scale, $f$.  Then, we simply assume the following ${\bf
Z}_2$ breaking soft mass terms,
\begin{align}
V_{\rm soft} &= \Delta \widetilde{m}_{H_u}^2 |H_u^A |^2
+\Delta \widetilde{m}_{H_d}^2 |H^A_d|^2 . \label{z2breakingmhu}
\end{align}

In order to make the discussion independent of the form of quartic
couplings, we take the decoupling limit of the SUSY heavy Higgses and
match the theory to the non-supersymmetric twin Higgs potential.  In the
decoupling limit, the four Higgs doublets can be written as follows in
terms of $H_A$ and $H_B$ in the non-supersymmetric twin Higgs
model,
\begin{equation}
\begin{split}
&H_u^{A} = H_A \sin \beta_{A}, \qquad H_u^{B}= H_B \sin \beta_{B},\\[1ex]
&H_d^{A} = H_A^{\dagger} \cos \beta_{A}, \qquad H_d^{B}= H_B^{\dagger} \cos \beta_{B}. \label{eq:decouple}
\end{split}
\end{equation}
Here, $\tan\beta_{A}= {v_u^{A}}/{v_d^{A}}$ and $\tan\beta_{B}=
{v_u^{B}}/{v_d^{B}}$ with $v_{u,d}^{A,B} \equiv \langle H_{u,d}^{A,B}
\rangle / \sqrt{2}$. Thanks to the approximate ${\bf Z}_2$
symmetry, they are almost equal, $\tan\beta_{A} \simeq \tan\beta_{B}$.
  In the rest of the discussion, we simply assume
$\beta_{A} = \beta_{B} = \beta$ and $\widehat{g}_2=g_2$. 
Note that by taking the decoupling limit \eqref{eq:decouple}, the structure of the Higgs potential is essentially the same as that of the potential given by \eqref{eq:general} discussed in the previous subsection.
When we require that the
supersymmetric twin Higgs potential is matched with
Eq.~\eqref{eq:general}, we obtain the following relations,
\begin{equation}
\begin{split}
-\lambda f^2 &= \widetilde{m}^2_{H_u} \sin^2 \beta + \widetilde{m}^2_{H_d}  \cos^2 \beta +\mu^2 -b\sin 2\beta , \\[1ex]
\sigma_1 f^2 &= (\sigma + \delta \sigma) f^2 = \Delta \widetilde{m}_{H_u}^2 \sin^2 \beta +\Delta \widetilde{m}_{H_d}^2 \cos^2 \beta +\delta \sigma f^2, \\[1ex]
\kappa_1 &= \kappa + \delta \kappa = \frac{g_2^2}{8} \cos^2 2\beta +\delta \kappa, \\[1ex]
\rho_1 &= \rho + \delta \rho = \frac{g_1^2}{8} \cos^2 2\beta +\delta \rho, \label{matchingrelation}
\end{split}
\end{equation}
where $\delta \sigma, \delta \kappa$ and $\delta \rho$ represent
the radiative corrections.
With these expressions, we can evaluate the SM-like Higgs mass
from \eqref{exacthiggsmass}.
Note that it is difficult to realize the SM-like Higgs mass only with the quartic couplings in the D-term potential, $\kappa = \frac{g_2^2}{8} \cos^2 2\beta$ and $\rho = \frac{g_1^2}{8} \cos^2 2\beta$.
We simply assume that there is an additional contribution or a radiative correction to $\kappa$ and $\rho$ to realize the SM-like Higgs mass.
As mentioned in the previous subsection, we impose the EWSB conditions and the conditions $\lambda > \sigma_1,~\kappa_1,~\rho_1$ to consider the general feature of SUSY twin Higgs models.
In Sec.~\ref{sec:twin phase} and Sec.~\ref{sec:U(4) breaking phase transition}, we will discuss the order of the phase transitions with imposing these conditions.

\section{The electroweak phase transition}\label{sec:twin phase}

As discussed in Sec.~\ref{sec:twin}, twin Higgs models generally
accommodate breakings of the two symmetries. One of them is the
standard EWSB and another is the breaking of
the $U(4)$ symmetry to the $U(3)$ one, through which the SM Higgs field
is identified with one of the pNGBs.  Let us call the phase transition
corresponding to the latter breaking the $U(4)$-breaking phase
transition. In this paper, we analyze not only the electroweak phase
transition but also this $U(4)$-breaking phase transition in cosmology.  In this
section, we study the order of the electroweak phase transition in twin Higgs models, with and
without SUSY, especially.

\begin{figure}[h]
\centering\includegraphics[width=8cm]{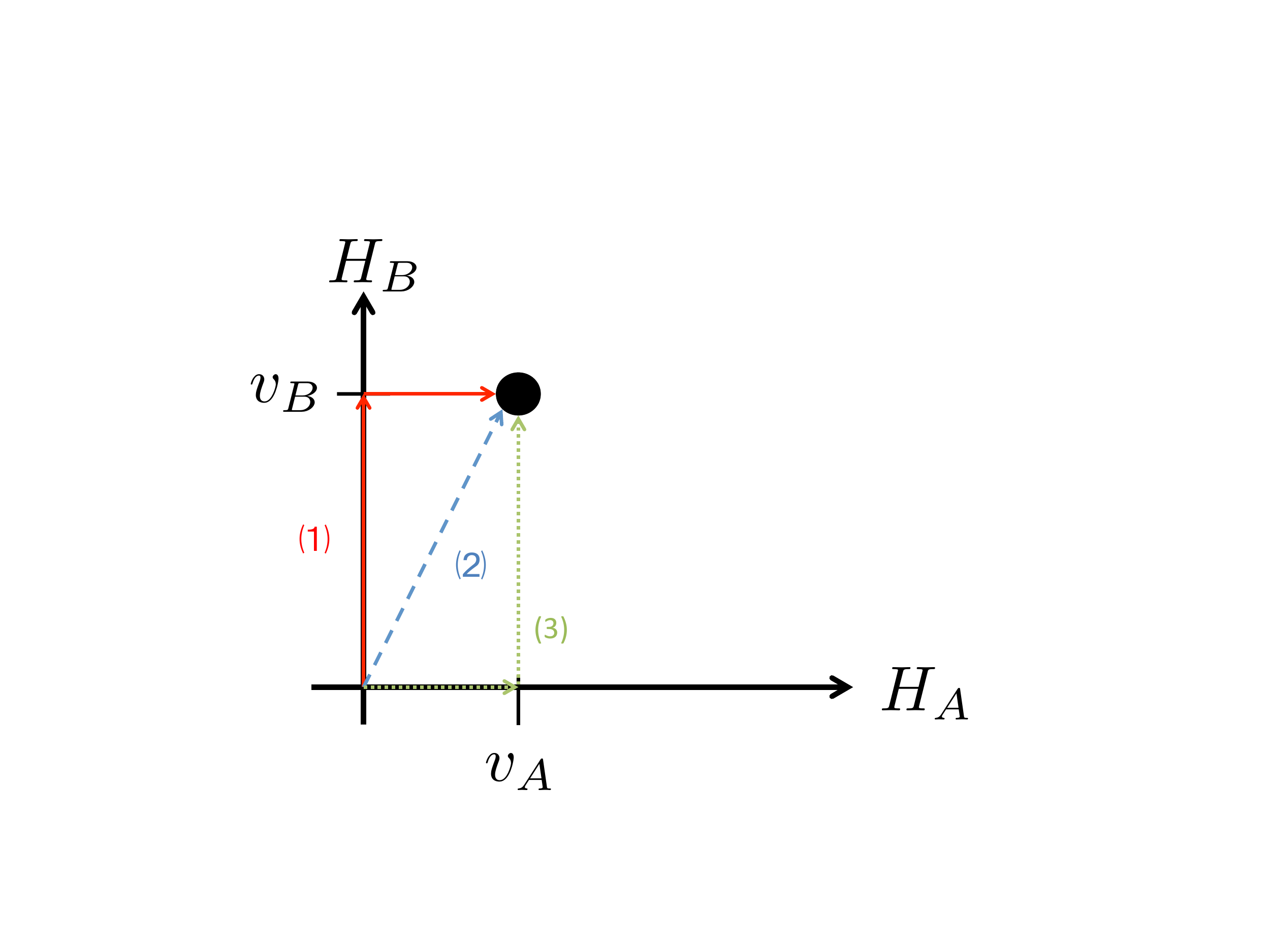}
\caption{This figure shows three possible trajectories of the phase
transitions.  In this figure, $v_A$ and $v_B$ are the vacuum expectation
values of the $H_A$ and $H_B$ fields at the zero temperature.  The black
point represents the true vacuum at the zero temperature.  We consider
only the path (1). }\label{fig:phase transition path}
\end{figure}

Before going to the detailed calculation, we would like to discuss the
thermal history in the early Universe.  At high-temperature, both of
$\phi_A$ and $\phi_B$ fields are trapped at the origin of the potential due to
the thermal mass terms.  When the temperature cools down, another minimum
different from the origin appears. Below the critical temperature, $\phi_A$
and $\phi_B$ fields eventually roll down or tunnel to the true vacuum,
and the $U(4)$ symmetry and its subgroup, the SM electroweak symmetry, finally break down.
However, we do not
know how these two phase transitions proceed.  Let us denote the
temperatures when $\phi_A$ and $\phi_B$ fields acquire their VEVs
by $T_A$ and $T_B$, respectively. In general, there are
three possible trajectories of these two phase transitions, which are
schematically described in Fig.~\ref{fig:phase transition path}.  The
red line (1) shows the trajectory of a two-step phase transition with
$T_B \gg T_A$, in which $\phi_B$ field acquires its VEV
first 
and $\phi_A$ field does later. The blue solid
line (2) shows the trajectory of a one-step phase transition with $T_A
\sim T_B$, in which the Higgs field rolls (or tunnels) to the true vacuum
directly. The green dotted line (3) shows the trajectory of another
two-step phase transition with $T_A \ll T_B$, in which $\phi_A$ field acquires its VEV first and 
$\phi_B$ field does later.
In this paper, we consider the case with $T_B \gg T_A$ 
and we call the phase transition at which $\phi_B$ field acquires its VEV, the $U(4)$-breaking 
phase transition.
Let us consider the condition under which this case happens.
The thermal mass terms for $\phi_A$ and $\phi_B$ fields are given by
\begin{align}
m_A^2 (H_A,~T)) &=(\zeta_A T^2 -(\lambda -\sigma_1 )f^2 ) |H_A|^2,\\
m_B^2 (H_B,~T)&= (\zeta_B T^2 -\lambda f^2) |H_B|^2, \label{eq:thermal massB}
\end{align} 
where $\zeta_A$ and $\zeta_B$ represent the numerical coefficients depending on the coupling constants.
The critical temperatures $T_A$ and $T_B$ are evaluated by the condition $m_A (T_A)=m_B (T_B)=0$, which yields
\begin{align}
\frac{T_A}{T_B} = \sqrt{\frac{\zeta_B}{\zeta_A}} \sqrt{1-\frac{\sigma_1}{\lambda}}.
\end{align}
Taking into account the twin ${\bf Z_2}$ symmetry $\zeta_A \simeq \zeta_B$, we obtain $T_A /T_B \simeq \sqrt{1-\sigma_1/ \lambda}$.
Therefore, $\sigma_1 >0$ is a necessary condition to realize $T_A \ll T_B$.
The region with $\sigma_1 >0 $ is also shown in Fig.~\ref{fig:allowed_region}.

We shall study the strength of the electroweak phase transition.
The thermal resummed effective
potential for both of $H_A$ and $H_B$ Higgs fields are calculated at the one-loop order in the same way as Ref.~\cite{Kilic:2015joa}. 
We take account of the top, twin top quarks, $SU(2)_W \times U(1)_Y$ gauge
bosons, and $SU(2)_{\widehat{W}}$ gauge bosons, respectively, because they
give dominant contributions to the effective potential.
The general expression of the thermal effective potential is summarized in appendix~\ref{sec:thermal effective potential}.

Let us calculate the thermal one-loop resummed effective potential
Eq.~\eqref{eq:ressumed effective potential} for $H_A$ and $H_B$ in the non-supersymmetric case
starting from the effective potential \eqref{eq:general}.  Since we
take account of not only the $H_A$ field but also the $H_B$ field, we
consider the following background fields,
\begin{align}
H_A =
\begin{pmatrix}
0\\
\dfrac{\phi_A}{\sqrt{2}}
\end{pmatrix},~H_B =
\begin{pmatrix}
0\\
\dfrac{\phi_B }{\sqrt{2}}
\end{pmatrix}.
\end{align}
The number of degrees of freedom (d.o.f) and the field dependent masses of $SU(2)_W
\times U(1)_Y$ gauge bosons, $SU(2)_{\widehat{W}}$ gauge bosons, top quark and twin top quark are
given by respectively
\begin{align}
&n_{W}=6,~~~m^2_{W}=\frac{g_2^2 \phi^2_A}{4},\\
&n_{Z}=3,~~~m_{Z}^2 =(g_1^2+g_2^2)\frac{\phi_A^2}{4} ,\\
&n_{\widehat{W}}={9},~~~m^2_{\widehat{W}}=\frac{\widehat{g}_2^2 \phi^2_B}{4},\label{eq:twin gauge bosons}\\
&n_{t}=12,~~~m^2_{t}= \frac{y_t^2 \phi_{A}^2}{2},\\
&n_{\widehat{t}}=12,~~~m^2_{\widehat{t}}=\frac{\widehat{y}_t^2 \phi_{B}^2}{2}.\label{eq:twin top}
\end{align}
Note that here we considered the Fraternal model where the mirror $U(1)$ gauge fields are absent, 
but that we expect that the basic results are unchanged even if we include them since the
$U(1)$ gauge coupling is tiny.
With the supersymmetric completions visible and mirror stops might also contribute, 
but we do not take account of them by assuming they are sufficiently heavy through the Higgs $\phi_B$'s VEV.
With this assumption, our conclusion is applicable also to the case with the supersymmetric UV completions.

The one-loop effective potential $V_{\rm eff}$ is then given by
\begin{align}
&V_{\rm eff} = V_{0} + V_{\rm CW} + V_{\rm thermal},\\
&V_0 (\phi_A,~\phi_B) =\frac{\lambda}{4}(\phi_A^2 +\phi_B^2 -f^2)^2 +\frac{\kappa}{4}(\phi_A^4 +\phi_B^4)+\frac{\rho}{4} \phi_A^4
+\frac{\sigma}{2} f^2 \phi_A^2 \label{eq:tree part}, \\
&V_{\rm CW} (\phi_A,~\phi_B)= 
- \frac{3}{16\pi^2}m_t^4 (\phi_A) \left(\log\left(\frac{m_t^2 (\phi_A)}{\mu^2}\right) -\frac{3}{2} \right)
- \frac{3}{16\pi^2}m_{\widehat{t}}^4 (\phi_B) \left(\log\left(\frac{m_{\widehat{t}}^2 (\phi_B)}{\mu^2}\right) -\frac{3}{2} \right)\nonumber\\
&~~~~~~~~~~~~~~~~~~~~~~~~+ \frac{3}{32\pi^2}m_W^4 (\phi_A) \left(\log\left(\frac{m_W^2 (\phi_A)}{\mu^2}\right) -\frac{3}{2} \right) + \frac{3}{64\pi^2}m_Z^4 (\phi_A) \left(\log\left(\frac{m_Z^2 (\phi_A)}{\mu^2}\right) -\frac{3}{2} \right)\nonumber\\
&~~~~~~~~~~~~~~~~~~~~~~~~+ \frac{9}{64\pi^2}m_{\widehat{W}}^4 (\phi_B) \left(\log\left(\frac{m_{\widehat{W}}^2 (\phi_B)}{\mu^2}\right) -\frac{3}{2} \right), \label{eq:zero temperature part} \\
&V_{\mathrm{thermal}} (\phi_A,~\phi_B,~T)= -\frac{6}{\pi^2}T^4 J_F \left[\frac{m_t^2 (\phi_A)}{T^2}\right]-\frac{6}{\pi^2}T^4 J_F \left[\frac{m_{\widehat{t}}^2 (\phi_B)}{T^2}\right]\nonumber \\
&~~~~~~~~~~~~~~~~~~~~~~~~+\frac{3}{\pi^2}T^4 J_B \left[\frac{m_W^2 (\phi_A)}{T^2}\right]+\frac{3}{2\pi^2}T^4 J_B \left[\frac{m_Z^2 (\phi_A)}{T^2}\right]
+\frac{9}{2\pi^2}T^4 J_B \left[\frac{m_{\widehat{W}}^2 (\phi_B)}{T^2}\right]. \label{eq:finite temperature part}
\end{align}
See the App.~\ref{sec:thermal effective potential} for the details.
Here, $\kappa,~\rho$ and $\sigma$ are the tree-level couplings and do not include the contribution of the one-loop Coleman-Weinberg potential.
In addition, we consider the ring diagram contributions denoted
by $V_{\mathrm{ring}}$ discussed in appendix~\ref{sec:thermal effective potential} to improve the perturbativity.
Since the masses of the $SU(2)_{\widehat{W}}$ gauge
bosons originating from the VEV of the $H_B$ field are much larger than
thermal corrections to the masses around the critical temperature, the
$SU(2)_{\widehat{W}}$ ring diagram contributions can be neglected. On the
other hand, the ring diagram
contributions coming from $SU(2)_W\times U(1)_Y$ gauge bosons are not negligible and
we need to take them into account. $V_{\mathrm{ring}}$ was computed in Ref.~\cite{Rose:2015lna} and is given by
\begin{align}
V_{\mathrm{ring}} =T\sum_{i=W_L,~Z_L,~\gamma_L} -\frac{n_i}{12\pi^2} \left((\overline{m}_i^2 (\phi_A,~T))^{\frac{3}{2}} - (m^2_i (\phi_A))^{\frac{3}{2}}\right),\label{eq:ring part}
\end{align}
with
\begin{align}
n_{W_L}=2,~~~\overline{m}_{W_L}^2 (\phi_A,~T) &= m_W^2 (\phi_A) + \frac{11}{6}g_2^2 T^2, \\
n_{Z_L}=1,~~~\overline{m}_{Z_L}^2 (\phi_A,~T)&= \frac{1}{2}\left[m_Z^2 (\phi_A)+\frac{11}{6}(g_2^2 +g_1^2)T^2 +\Delta(\phi_A,T) \right],\\
n_{\gamma_L}=1,~~~\overline{m}_{\gamma_L}^2 (\phi_A,~T)&= \frac{1}{2}\left[m_Z^2 (\phi_A)+\frac{11}{6}(g_2^2 +g_1^2)T^2 -\Delta(\phi_A,T) \right],\\
\Delta &=\sqrt{m^4_Z (\phi_A)+\frac{11}{3} \frac{(g_2^2 -g_1^2 )^2}{g_2^2+g_1^2} \left[m_{Z}^2+\frac{11}{12}(g_2^2 +g_1^2)T^2\right]T^2}.
\end{align}
Here, $n_i$ represents the number of d.o.f. for each longitudinal mode.
We do not take account of transverse modes of the $SU(2)_{W} \times U(1)_Y$ gauge bosons 
because the magnetic masses they receive from the environment are suppressed by $g_1^4$ or $g_2^4$ and 
hence give minor contributions.

In our case, the electroweak phase transition occurs after the $U(4)$-breaking phase transition. Therefore, during the electroweak phase
transition, $\phi_B$ already gets a non-zero VEV, $\phi_B (T)\neq 0$.
Then, in the same way as Eq.~\eqref{eq:EFT}, we integrate out the $\phi_B(T)$
field by setting
\begin{align}
\phi_B^2 (T)= f^2 -\phi_A^2 (T). \label{eq:finite EFT}
\end{align}
Here, $\phi_{A(B)} (T)$ represent the temperature dependent VEVs,
respectively. It should be noticed that, when we take the $T=0$ limit,
Eq.~\eqref{eq:finite EFT} is reduced to Eq.~\eqref{eq:EFT}. The one-loop
resummed effective potential Eq.~\eqref{eq:ressumed effective potential} for
$\phi_A$ can be written as
\begin{align}
V(\phi_A,~T) = V_0(\phi_A,~f^2-\phi_A^2)+V_{\rm CW}(\phi_A,~f^2-\phi_A^2)
+V_{\mathrm{thermal}}(\phi_A,~f^2-\phi_A^2,~T)+V_{\mathrm{ring}}(\phi_A,~T), \label{eq:finite EFT potential}
\end{align}
where $V_0,~V_{\rm CW},~V_{\mathrm{thermal}}$, and $V_{\mathrm{ring}}$ are
given by Eq.~\eqref{eq:tree part},~\eqref{eq:zero temperature
part},~\eqref{eq:finite temperature part}, and~\eqref{eq:ring part},
respectively. For the zero temperature part $V_0+V_{\rm CW}$, we set the
renormalization conditions given by
\begin{align}
&\left.\frac{d}{d\phi} (V_0+ V_{\rm CW})\right|_{\phi=v_A} = 0, \label{ren1}\\
&\left.\frac{d^2}{d\phi^2} (V_0 +V_{\rm CW})\right|_{\phi=v_A} =2\lambda_{SM} v_A^2.\label{ren2}
\end{align}
Neglecting $\mathcal{O}(\phi_A^6) $ terms, we obtain the following expression,
\begin{align}
V_0 +V_{\rm CW} &= -\frac{\lambda_{SM}}{2}v_A^2 \phi_A^2 +\frac{\lambda_{SM}}{4}\phi^4_A \nonumber \\
&+ \frac{n_i}{64\pi^2}\sum_i \left(m_i^4 (\phi_A)\left(\log\left( \frac{m_i^2 (\phi_A)}{m_i^2(v_A)}\right)-\frac{3}{2}\right)+2m_i^2 (v_A) m_i^2 (\phi_A) \right), \label{eq:renormalized potential}
\end{align}
where the suffix $i$ represents only the SM contribution. 
Now the system is parameterized only by the $U(4)$-breaking scale $f$ since 
the condition \eqref{eq:finite EFT} and renormalization condition 
Eqs.~\eqref{ren1} and \eqref{ren2} completely fix the other model parameters, 
$\kappa_1, \sigma_1,\rho_1$, and $\lambda$.

In Ref.~\cite{Kilic:2015joa}, it was shown that the one-loop effective
potential obtained by use of the relation Eq.~\eqref{eq:finite EFT} exhibits 
the restoration of the electroweak symmetry at high temperature, which
guarantees the presence of the electroweak phase transition. In this
paper, we try to clarify the order of the electroweak phase transition.
For this purpose, we will first check the validity of perturbative expansion near the
critical temperature. As is seen in App.~\ref{app:finite temperature},
the perturbative expansion is valid only when the following condition is satisfied,
\begin{align}
g_2^2 \frac{T_C}{m_W (\phi_A (T_C))} \sim g_2 \frac{T_C}{\phi_A(T_C)}<1, 
\label{eq:perturbative criteria}
\end{align}
where $T_C$ is the critical temperature of the EWSB. 
\begin{figure}[h]
\centering\includegraphics[width=9cm]{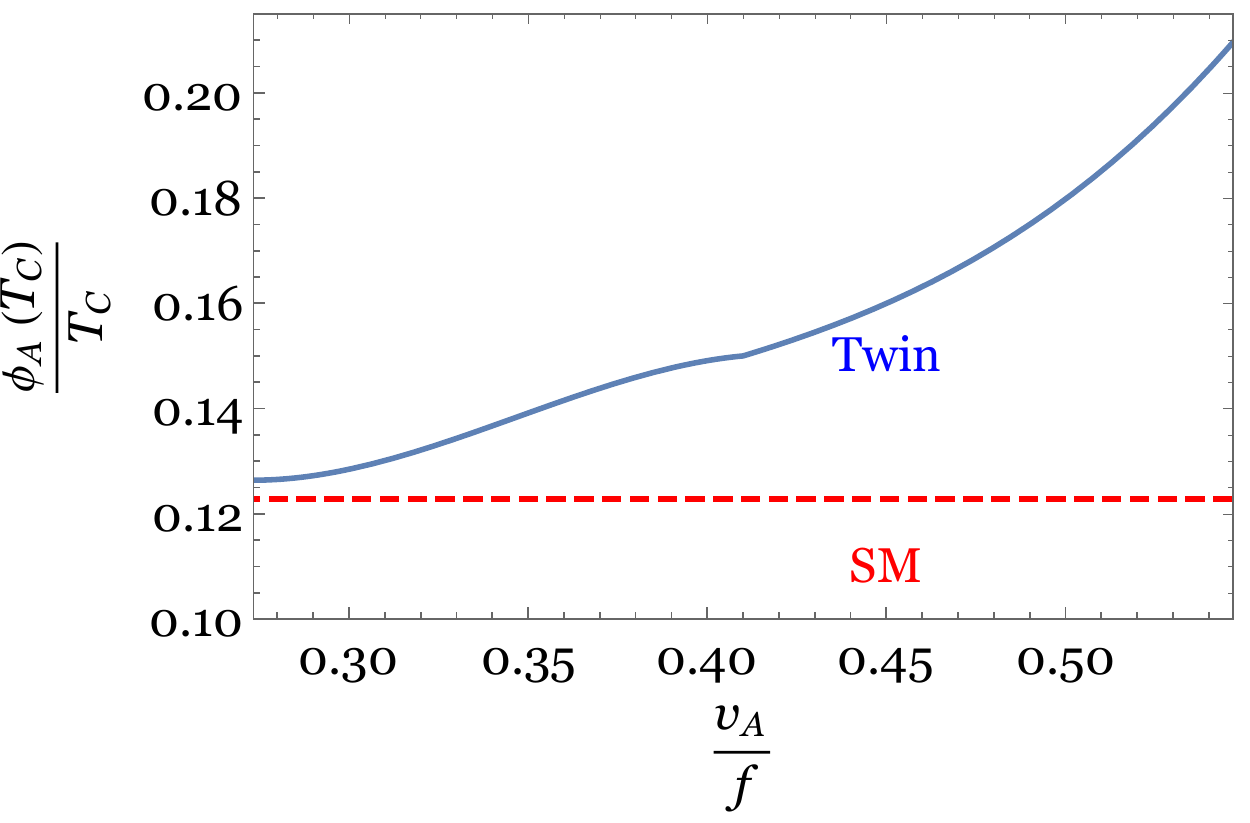}
\caption{The ratio $\phi_A(T_C) /T_C$ for each $U(4)$ symmetry breaking scale $f$.
The blue curve represents $\phi_A(T_C)/T_C$ evaluated
by use of Eq.~\eqref{eq:finite EFT potential}. The dashed red line represents
the same ratio but with only the Standard Model contributions being
taken into account.}\label{Fig1}
\end{figure}
Here the critical temperature $T_C$ is defined so that the electroweak symmetry preserving and breaking vacua
are degenerate. $\phi_A (T_C)$ represents the expectation value of $\phi_A$
for a breaking phase at $T_C$.
In Fig.~\ref{Fig1}, the ratio $\phi_A (T_C) /T_C$ is plotted for
each $U(4)$ symmetry breaking scale $f$.
We have evaluated the one-loop resummed effective potential given in
Eq.~\eqref{eq:finite EFT potential} without resort to the high temperature
expansions. It is easily seen that the larger a breaking scale $f$ is,
the smaller $\phi_A (T_C)/T_C$ is.  This fact can be easily understood as
follows. The thermal contributions from the twin particles could 
strengthen the first order nature of the EWSB. However, the twin partners acquire masses
proportional to $\phi_B (T)$ through the Higgs mechanism. Thus, a larger
breaking scale $f$ leads to larger masses of the twin particles, which
easily induces thermal decoupling of twin particles during the
electroweak phase transition. This decoupling makes $\phi_A (T_C)/T_C$ in our
case approach the value in the standard model case. 
Hence a larger $v_A/f$ indeed increases 
$\phi_A (T_C)/T_C$. However, the largest value of $\phi_A (T_C)/T_C$ for $f > 2v_A$ required by the constraint of the Higgs coupling measurement is at most $0.2$, which is not large enough to satisfy the criteria \eqref{eq:perturbative criteria}.
Therefore, we conclude that the higher order effects cannot be 
neglected  and the perturbative expansion is not
valid near $T_C$.
For the correct analysis, lattice simulations are required.
This result has an important implication for the electroweak
baryogenesis because it requires the sphaleron decoupling condition
$\phi_A (T_C) / T_C > 1$ around the critical temperature. Our results strongly suggest that this condition
is hardly satisfied in the Fraternal twin Higgs model 
as long as the condition \eqref{eq:finite EFT} is valid, 
and we cannot expect for the implementation of the electroweak baryogenesis.
This conclusion remains unchanged even if we go beyond the Fraternal model, as long as the condition \eqref{eq:finite EFT} and the assumption of the trajectory of two-step phase transition are adopted.
We do not exclude the possibility to have the strong first order electroweak phase transition 
once we relax one of these assumptions, which is beyond the scope of the present study.

Finally let us comment on some issues on UV completions.
We here do not assume concrete UV physics (SUSY and composite Higgs) in
our analysis and analyze the electroweak phase transition by use
of effective field theory for the $\phi_A$ field. 
As long as its usage is valid, our result is still robust
in supersymmetric and composite twin Higgs models.
However, it was shown in
Refs.~\cite{Bruggisser:2018mus,Bruggisser:2018mrt} that the electroweak
phase transition can be the strong first order in the composite Higgs scenario.
In the setup adopted in Refs.~\cite{Bruggisser:2018mus,Bruggisser:2018mrt}, the electroweak phase transition and the confinement
phase transition, which corresponds to the $U(4)$-breaking phase
transition in twin Higgs models, occurred simultaneously.  In addition,
the SM-like Higgs field couples with an additional scalar field.  
Thus this approach does not apply to our consideration.

\section{The $U(4)$-breaking phase transition} \label{sec:U(4) breaking phase transition}

In this section, we explore the $U(4)$-breaking phase transition in twin Higgs models with and without supersymmetric completion.
For the concrete calculation, we adopt the Fraternal model, but general features
would apply to other models.\footnote{See Ref.~\cite{Croon:2018erz} for general discussions of gravitational wave productions from a first order phase transition associated with $SU(N)$ breaking into $SU(N-1)$ in a hidden sector.}
As discussed in Sec.~\ref{sec:twin phase}, we assume that the $U(4)$-breaking phase transition occurs first and the electroweak phase transition does next in the following discussion.

\subsection{The case of the twin Higgs model without UV completion}\label{sec:minimal}

Let us first consider the twin Higgs model without any UV completions, in the sense 
that no new particles other than the mirror particles to the SM are involved.
The $U(4)$-breaking phase transition generally depends on UV physics such as SUSY and composite Higgs. However,
if new particles in the UV completion are sufficiently heavy during the phase transition, we can safely neglect the effect of these particles.
We shall study the strength of the $U(4)$-breaking phase transition by using the potential \eqref{eq:general} with this assumption. 

In our set up, the Higgs field $H_A$ is trapped at the origin of the potential $H_A =0$ due to the thermal mass term during the $U(4)$-breaking phase transition.
Thus, we take the background fields as
\begin{align}
H_A=
\begin{pmatrix}
0\\
0\end{pmatrix},\qquad H_B=
\begin{pmatrix}
0\\
\dfrac{\phi_B}{\sqrt{2}}
\end{pmatrix} . \label{eq:so8 background}
\end{align}
and calculate the resummed one-loop potential given by Eq.~\eqref{eq:ressumed effective potential} for the field $H_B$.
We take account of the twin top and $SU(2)_{\widehat{W}}$ gauge bosons which give 
dominant contributions to the effective potential.
On the other hand, a larger quartic coupling $\lambda > \widehat{g}^2_2$ makes the $U(4)$-breaking phase transition weaker $\phi_B (T_C) / T_C < \widehat{g}_2$ (see Eq.~\eqref{eq:strong first order}) hence we consider a small quartic coupling $\lambda < \widehat{g}^2_2$.
Since the quartic coupling $\lambda < \widehat{g}_2^2$ is smaller than $\widehat{g}_2$ and $y_{\widehat{t}}$, we neglect the $H_A$ and $H_B$ loop contributions to the effective potential in the following discussion.

With the field dependent masses of the twin top and $SU(2)_{\widehat{W}}$ gauge bosons given by Eqs.~\eqref{eq:twin gauge bosons} and~\eqref{eq:twin top},
The one-loop effective potential $V_{\rm eff}$
is expressed as
\begin{align}
&V_{\rm eff}=V_{0} +V_{\rm CW} + V_{\rm thermal},\\
&V_0 (\phi_B) = \frac{\lambda}{4} (\phi_B^2-f^2)^2 +\frac{\kappa}{4}\phi_B^4 = -\frac{\lambda}{2} f^2 \phi_B^2 +\frac{\lambda+ \kappa}{4} \phi_B^4 +\frac{\lambda}{4}f^4, \\
&V_{\rm CW} (\phi_B) = -\frac{3}{16\pi^2}m_{\widehat{t}}^4 (\phi_B) \left(\log\left( \frac{m^2_{\widehat{t}}(\phi_B)}{\mu^2}\right)-\frac{3}{2}\right)+\frac{9}{64\pi^2}m_{\widehat{W}}^4 (\phi_B) \left(\log\left( \frac{m^2_{\widehat{W}} (\phi_B)}{\mu^2}\right)-\frac{3}{2}\right), \\
&V_{\mathrm{thermal}} (\phi_B,~T) = -\frac{6}{\pi^2} T^4 J_F \left[\frac{m_{\widehat{t}}^2 (\phi_B)}{T^2}\right]
+\frac{9}{2\pi^2}T^4 J_B \left[\frac{m_{\widehat{W}}^2 (\phi_B)}{T^2}\right] \label{eq:thermal so8}.
\end{align}
In order to find the ring diagram contribution $V_{\rm ring}$, we need to evaluate the thermal masses of the $SU(2)_{\widehat{W}}$ gauge bosons denoted by \eqref{eq:thermal field dependent mass}. 
We here take account of the one-loop self-energy of longitudinal modes~\cite{Comelli:1996vm}
\begin{align}
\Pi_{\widehat{W}_L} &= \frac{7}{6}\widehat{g}_2^2 T^2,  \label{thermalmass}
\end{align} 
in which only one generation (third generation) is included for the Fraternal model. 
The transverse modes receive the magnetic masses, but they are suppressed by the factor of ${\widehat g}^4_2$ and 
hence we omit them as is the case of the electroweak phase transition.
We then obtain the ring diagram contribution $V_{\mathrm{ring}}$ given by
\begin{align}
V_{\mathrm{ring}} &= -\frac{T}{4\pi} \left(\left(\overline{m}^{2}_{\widehat{W}_L} (\phi_B,~T)\right)^{\frac{3}{2}} -\left(m^2_{\widehat{W}} (\phi_B)\right)^{\frac{3}{2}} \right),\\
\overline{m}^2_{\widehat{W}_L} &= m_{\widehat{W}}^2 (\phi_B) +\Pi_{\widehat{W}_L} .
\end{align}
When we use the high-temperature expansions given by Eq.~\eqref{eq:boson} and Eq.~\eqref{eq:fermion}, the resummed one-loop effective potential takes the following form:
\begin{align}
V&= V_0 +V_{\rm CW}+V_{\mathrm{Thermal}}+V_{\mathrm{ring}}\nonumber\\
&=\frac{1}{2}M^2(T)\phi_B^2-\frac{T}{2\pi}\left(\frac{\widehat{g}_2^2 \phi_B^2}{4}\right)^{3/2}
 -\frac{T}{4\pi} \left(\frac{ \widehat{g}_2^2 \phi_B^2 }{4}+\Pi_{\widehat{W}_L}  \right)^\frac{3}{2}+\frac{\lambda+\kappa_1 (T) }{4}\phi_B^4, \label{eq:so8 potential}
 \end{align}
where
\begin{align}
 &M^2(T)=-\lambda f^2+\frac{\widehat{y}_{t}^2}{4}T^2+\frac{3\widehat{g}_2^2}{16}T^2, \\
 &\kappa_1(T)=\kappa-\frac{3\widehat{y}_t^4}{16\pi^2}\left(\log\left(\frac{a_f T^2}{\mu^2}\right)-\frac{3}{2}\right)+\frac{9\widehat{g}_2^4}{256\pi^2}\log\left(\left(\frac{a_b T^2}{\mu^2}\right)-\frac{3}{2}\right) \label{eq:self-coupling correction}.
\end{align}
Thanks to the twin ${\bf Z_2}$ symmetry, $y_t\simeq \widehat{y}_t$ and $g_2\simeq \widehat{g}_2$, the $U(4)$-breaking phase transition described by the potential \eqref{eq:so8 potential} is similar to the electroweak phase transition in the SM which
has been analyzed by perturbative~\cite{Dine:1992wr} and non-perturbative~\cite{Csikor:1998eu,Rummukainen:1998as} methods.
The most reliable approach to clarify the order of the phase transition is the lattice simulation~\cite{Csikor:1998eu,Rummukainen:1998as}.
It was shown that the electroweak phase transition in the SM is the first order
when $m_H\lesssim 70-80$ GeV (or $\lambda_{\rm SM}\lesssim 0.04$) is satisfied.

One might wonder if the difference between $U(4)$-breaking sector and the SM sector 
prevents us from adopting the results of the SM to the $U(4)$-breaking case. 
But these differences are negligible for our purpose in the following reasons.
First of all, the breaking scale of the $U(4)$-breaking phase transition, $f$, is different from that of the electroweak phase transition, $v_A$.  
However, the order of the electroweak phase transition in the SM depends on the parameter $\lambda_{\rm SM} / g_2^2$
\cite{Rummukainen:1998as}, but not $v_A$.
Thus, we have only to identify $\lambda + \kappa_1$ in our model with $\lambda_{\rm SM}$ in the 
SM electroweak phase transition.
Second, there is no $U(1)_{\widehat{Y}}$ gauge boson in the Fraternal twin Higgs model.
Since the $U(1)_{\widehat{Y}}$ gauge coupling $\widehat{g}_1$ is tiny compared to the $SU(2)_{\widehat{W}}$ gauge coupling $\widehat{g}_2$, we can neglect this effect safely. 
Indeed, the original paper \cite{Rummukainen:1998as} also does not include 
$U(1)_Y$ and they concluded that the error due to this assumption is small enough.
Finally, the coefficient of the thermal mass \eqref{thermalmass} for the Fraternal model\footnote{In the Mirror twin Higgs models the coefficient of the thermal mass is the same to the SM.}  differs from that of the $SU(2)_{W}$ gauge bosons in the SM.
In the lattice simulation~\cite{Rummukainen:1998as}, they use the three-dimensional effective Lagrangian obtained by integrating out all fermions and the longitudinal modes of $SU(2)_W$ gauge bosons~\cite{Farakos:1994kx,Kajantie:1995dw,Kajantie:1997ky}, which affects
the values of the parameters $\lambda_{\rm SM}$ and $g_2$.
We confirmed that this difference gives only 10 percent changes in the parameters of three-dimensional effective Lagrangian, and hence we can safely neglect it.
Therefore, the order of the $U(4)$-breaking phase transition can be analyzed
by use of the result of the electroweak phase transition in the SM.
We conclude that the $U(4)$-breaking phase transition is the first order when $\lambda + \kappa_1 \lesssim 0.04$ is satisfied, thanks to $y_t \simeq y_{\widehat{t}}$ and $g_2 \simeq \widehat{g}_{2}$.

As discussed in Sec.~\ref{sec:minimal twin Higgs}, the parameters $\kappa_1$ and $\lambda$ are bounded below, $\lambda + \kappa_1  \gtrsim 0.1$,  due to the EWSB conditions and the conditions $\lambda > \sigma_1,~\kappa_1,~\rho_1$ as we can see from Fig.~\ref{fig:allowed_region}.
Therefore, it cannot satisfy the condition for the first order $U(4)$-breaking phase transition, $\lambda + \kappa_1 \lesssim 0.04$.
We also expect no gravitational wave production because of the absence of a first order phase transition in the case of twin Higgs models without any UV completions. 
The differences in the Fraternal and Mirror models give minor effects 
and are within the uncertainties in our estimate.  More generally, our conclusion is 
robust in any models as long as the tree-level potential is given by Eq.~\eqref{eq:general}
and there are no additional light degrees of freedom so that the thermal masses for twin $SU(2)_{\widehat{W}}$ gauge bosons do not differ so much.

\subsection{The case of supersymmetric twin Higgs models}\label{sec:transition with twin stop}

In the previous subsection, we do not consider effects of UV physics
such as composite Higgs and SUSY on the $U(4)$-breaking phase transition.
If other fields strongly couple to the Higgs field $H_B$, we cannot apply the argument in the previous subsection.
We here consider supersymmetric twin Higgs models and explore the order of the $U(4)$-breaking phase transition.
Especially, since any such models contain twin stops, 
which are strongly coupled to the Higgs field $H_B$ 
and possibly light at the restored phase in the absence of the Higgs VEV, 
we focus on the effect of light twin stops. 
Hereafter, we take the decoupling limit, simply assuming that every supersymmetric partner except for twin stops acquires a large soft mass and
decouples with thermal plasma during the $U(4)$-breaking phase transition.
We will show that there is some parameter space where the $U(4)$-breaking phase transition
is the first order and estimate the gravitational wave amplitude generated through this phase transition.

Let us calculate the one-loop resummed effective potential (\ref{eq:ressumed effective potential}).
Since we take the decoupling limit as explained in section \ref{sec:SUSY twin Higgs}, the background fields are given by
\begin{align}
H^A_u =
\begin{pmatrix}
0\\
0
\end{pmatrix},~~H^A_d =
\begin{pmatrix}
0\\
0
\end{pmatrix},~~
H_u^B= 
\begin{pmatrix}
0\\
\dfrac{\phi_B}{\sqrt{2}}\sin \beta
\end{pmatrix},~~H^B_d=
\begin{pmatrix}
\dfrac{\phi_B}{\sqrt{2}}\cos \beta\\
0
\end{pmatrix}.
\end{align}
We take account of the left and right-handed twin stops, the twin top quarks and the 
$SU(2)_{\widehat{W}}$ gauge bosons, which give dominant contributions to the effective potential.
We neglect the Higgs loop correction as is the non-supersymmetric case.
The tree level potential $V_0$ in Eq.~\eqref{eq:ressumed effective potential} is  concretely written as
\begin{align}
V_0 = \frac{\lambda}{4} (\phi_B^2 -f^2)^2 +\frac{\kappa}{4}\phi_B^4 , \label{eq:supersymmetric U(4) potential}
\end{align}
where the second term includes the $D$-term contribution, $\kappa \supset \widehat{g}_2^2 \cos^2 2\beta / 8 $. 
The thermal one-loop corrections $V_{\rm CW}$ and $V_{\rm thermal}$ are evaluated as follows.
The field dependent masses of twin top quarks and $SU(2)_{\widehat{W}}$ gauge bosons are given by Eqs.~\eqref{eq:twin top} and \eqref{eq:twin gauge bosons}.
Those of the left and right-handed twin stops can be written as
\begin{align}
 \mathcal{M}^2_{\widehat{\mathrm{stop}}}=
 \begin{pmatrix}
 \widetilde{m}_{\widehat{Q}}^2+m_{\widehat{t}}^2 (\phi_B)+\frac{\widehat{g}_2^2}{8}\phi_B^2 \cos 2 \beta & m_{\widehat{t}}(\phi_B)X_{\widehat{t}}\\
 m_{\widehat{t}}(\phi_B) X_{\widehat{t}} & \widetilde{m}_{\widehat{t}_R}^2  + m_{\widehat{t}}^2 (\phi_B)
 \end{pmatrix},~~~X_{\widehat{t}} \equiv A_{\widehat{t}} - \mu \cot \beta, \label{eq:mass matrix}
\end{align}
where $\widetilde{m}_{\widehat{Q}}^2,~\widetilde{m}_{\widehat{t}_R}^2$ and $A_{\widehat{t}}$ are the twin left,~right-handed stop soft mass-squared and the twin A-term, respectively.
The diagonalized masses are given by
\begin{align}
n^{1(2)}_{\widetilde{t}} &=6, \\
m^2_{\tilde{t}_{1,2}^B} (\phi_B) &= \frac{(\mathcal{M}^2_{\widehat{\mathrm{stop}}})_{11}+(\mathcal{M}^{2}_{\widehat{\mathrm{stop}}})_{22} }{2}\pm \sqrt{\left(\frac{(\mathcal{M}^2_{\widehat{\mathrm{stop}}})_{11}-(\mathcal{M}^2_{\widehat{\mathrm{stop}}})_{22} }{2}\right)^2 +\left((\mathcal{M}_{\widehat{\mathrm{stop}}}^{2} )_{12} \right)^2 }~,
\end{align}
where $n^{1(2)}_{\widetilde{t}}$ and the superscript of $\mathcal{M}_{\widehat{\rm stop}}^2$ represent the number of d.o.f for the left (or right)-handed twin stop and the component of $\mathcal{M}_{\widehat{\mathrm{stop}}}^2$ matrix, respectively.
The one-loop effective potential is then written as
\begin{align}
V_{\rm CW} (\phi_B)&= -\frac{3}{16\pi^2}m_{\widehat{t}}^4 (\phi_B) \left(\log\left( \frac{m^2_{\widehat{t}}(\phi_B)}{\mu^2}\right)-\frac{3}{2}\right)+\frac{9}{64\pi^2}m_{\widehat{W}}^4 (\phi_B) \left(\log\left( \frac{m^2_{\widehat{W}}(\phi_B)}{\mu^2}\right)-\frac{3}{2}\right) \nonumber\\
&+\frac{3}{32\pi^2}m_{\tilde{t}_{1}^B}^4 (\phi_B) \left(\log\left( \frac{m_{\tilde{t}_{1}^B}^2 (\phi_B)}{\mu^2}\right)-\frac{3}{2}\right)
+\frac{3}{32\pi^2}m_{\tilde{t}_{2}^B}^4 (\phi_B) \left(\log\left( \frac{m_{\tilde{t}_{2}^B}^2 (\phi_B)}{\mu^2}\right)-\frac{3}{2}\right), \\
V_{\mathrm{thermal}} (\phi_B,~T) &= -\frac{6}{\pi^2} T^4 J_F \left[\frac{m_{\widehat{t}}^2 (\phi_B)}{T^2}\right]
+\frac{9}{2\pi^2}T^4 J_B \left[\frac{m_{\widehat{W}}^2 (\phi_B)}{T^2}\right] \nonumber \\
&+\frac{3}{\pi^2}T^4 J_B \left[\frac{m_{\tilde{t}_{1}^B}^2 (\phi_B)}{T^2}\right]
+\frac{3}{\pi^2}T^4 J_B \left[\frac{m_{\tilde{t}_{2}^B}^2 (\phi_B)}{T^2}\right].
\end{align}

In order to calculate the ring diagram contribution $V_{\rm ring}$, we need to evaluate thermal masses of the longitudinal mode of the $SU(2)_{\widehat{W}}$ gauge bosons.
As a result of the twin ${\bf Z}_2$ symmetry, thermal masses of the $SU(2)_{\widehat{W}}$ gauge bosons can be calculated in the same way as the case of the MSSM~\cite{Comelli:1996vm}.
The thermal masses of the longitudinal mode of the $SU(2)_{\widehat{W}}$ gauge bosons, the left and right-handed twin stops are given by
\begin{align}
\Pi_{\widehat{W}_L}&= \frac{5\widehat{g}_2^2}{3}T^2, \\
\Pi_{\widehat{t}_{R}}&=\frac{4}{9}\widehat{g}_3^2 T^2 + \frac{\widehat{y}_t^2}{6}\left(1+\frac{1}{\sin^2\beta}\right)T^2,\\ 
\Pi_{\widehat{Q}}&=\frac{4}{9}\widehat{g}_3^2 T^2 + \frac{\widehat{y}_t^2}{12}\left(1+\frac{1}{\sin^2\beta}\right)T^2 + \frac{\widehat{g}_2^2 }{4}T^2.
\end{align}
Then the temperature dependent mass matrix is given by
\begin{align}
\overline{\mathcal{M}}^2_{\widehat{\mathrm{stop}}}&=
 \begin{pmatrix}
 \widetilde{m}_{\widehat{{Q}}}^2+m_{\widehat{t}}^2 (\phi_B)+\frac{\widehat{g}_2^2}{8}\phi_B^2 \cos 2 \beta + \Pi_{\widehat{Q}} & m_{\widehat{t}}(\phi_B) X_{\widehat{t}}\\
 m_{\widehat{t}}(\phi_B) X_{\widehat{t}} & \widetilde{m}_{\widehat{t}_R}^2  + m_{\widehat{t}}^2 (\phi_B)+\Pi_{\widehat{t}_R}
 \end{pmatrix},\\
n^{1(2)}_{\widetilde{t}} &=6 ,
\end{align}
where $n^{1(2)}_{\widetilde{t}}$ represents the number of d.o.f for the left (or right)-handed twin stop.
From this expression, the ring diagram contributions are calculated as follows.
\begin{align}
V_{\mathrm{ring}} =& -\frac{T}{4\pi} \left((\overline{m}^{2}_{\widehat{W}_L} (\phi_B,~T))^{\frac{3}{2}} -(m^2_{\widehat{W}} (\phi_B) )^{\frac{3}{2}} \right)\nonumber \\
&-\frac{T}{2\pi}\left( ( \overline{m}^2_{\tilde{t}_{1}^B} (\phi_B))^{\frac{3}{2}} - ( m^2_{\tilde{t}_{1}^B} (\phi_B)  )^{\frac{3}{2}}+( (\overline{m}^2_{\tilde{t}_{2}^B} (\phi_B) )^{\frac{3}{2}}- ( m^2_{\tilde{t}_{2}^B} (\phi_B)  )^{\frac{3}{2}} \right),~\\
\overline{m}^2_{\tilde{t}_{1,2}^B} (\phi_B) &= \frac{(\overline{\mathcal{M}}^{2}_{\widehat{\mathrm{stop}}})_{11}+(\overline{\mathcal{M}}^{2}_{\widehat{\mathrm{stop}}})_{22}}{2} \pm \sqrt{\left(\frac{(\overline{\mathcal{M}}_{\widehat{\mathrm{stop}}}^{2})_{11}-(\overline{\mathcal{M}}_{\widehat{\mathrm{stop}}}^{2})_{22} }{2}\right)^2 + \left( (\overline{\mathcal{M}}_{\widehat{\mathrm{stop}}}^{2})_{12}\right)^2 } .
\end{align}

Moreover, in our set up, the twin QCD two-loop contribution is non-negligible compared to the resummed one-loop effective potential because the strong coupling $\widehat{g}_3$ and the top Yukawa coupling $\widehat{y}_t$ are large compared to the other matter couplings.
In the MSSM, the sunset diagram, which gives the dominant contribution, is evaluated in Ref.~\cite{Espinosa:1996qw}.
We adopt it and calculate the two-loop twin QCD contribution as 
\begin{align}
V^{(2)}_{\mathrm{thermal}} = - \frac{\widehat{g}_3^2}{2\pi^2}T^2\left((\overline{m}^2_{\tilde{t}_{1}^B} (\phi_B))^2 \log\left(\frac{2 \overline{m}^2_{\tilde{t}_{1}^B} (\phi_B)}{3T}\right)+(\overline{m}^2_{\tilde{t}_{2}^B} (\phi_B))^2 \log\left(\frac{2\overline{m}^2_{\tilde{t}_{2}^B} (\phi_B)}{3T}\right) \right). \label{eq:twin gluon two-loop}
\end{align}
In this expression, the high-temperature expansion~\cite{Parwani:1991gq} and mass-averaging approximation~\cite{Arnold:1992rz} are used. According to the discussions in Refs.~\cite{Funakubo:2012qc,Laine:2000kv}, their usage is justified for our purpose.
It should be noticed that this negative logarithmic dependence of $\phi_B$ in Eq.~\eqref{eq:twin gluon two-loop} gives an additional contribution to the potential barrier between the origin and another minimum.
Without taking this contribution into account, we would underestimate $\phi_B (T_C) / T_C$.

\begin{figure}[h]
\centering\includegraphics[width=9cm]{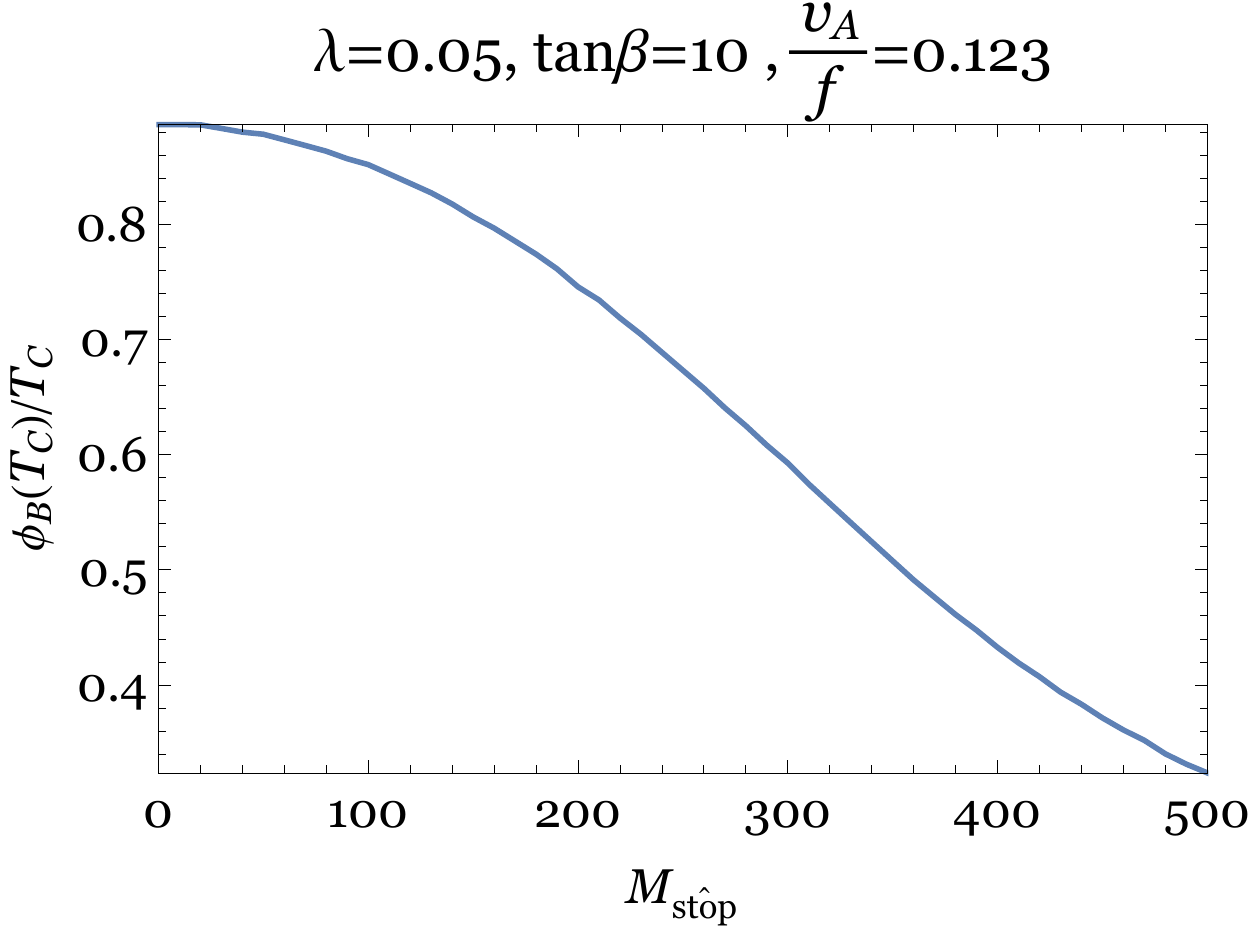}
\caption{This graph shows $\phi_B (T_C) /T_C$ and common left, right-handed twin stop soft masses. We set some physical parameters as $\lambda=0.05,~\widehat{g}_3 =1,~\tan\beta=10,~X_{\widehat{t}}=0$ and $v_A / f=0.123$.}\label{fig:firstordersoftmass}
\end{figure}

As discussed in App.~\ref{sec:thermal effective potential}, in order to have the first order phase transition and gravitational wave production,
$\phi_B (T_C) / T_C \gtrsim {\widehat g}_2$ is required. 
This ratio gets larger for a smaller $\lambda + \kappa_1$ (see App.~\ref{sec:thermal effective potential}.). 
As discussed in Sec.~\ref{sec:twin} (see Fig.~\ref{fig:allowed_region}), 
we have the conditions $\lambda>0.05$ and $\kappa_1>0.05$,  
from the requirements $\lambda > \rho_1, \kappa_1$ and $m_h\simeq 125$ GeV. 
Thus hereafter we take $\lambda \simeq 0.05$ and $\kappa_1 \simeq 0.05$ as the benchmark point. 
For simplicity we require the quartic coupling $\kappa_1$ is dominated by the D-term, $\kappa \simeq (\widehat{g}_2^2/8) \cos^2 2\beta$, 
so that  $\tan \beta \simeq 10$. 
The value of the twin QCD coupling constant $\widehat{g}_3$ can be somewhat different from the value of the visible QCD coupling constant $g_3 $ because the exact ${\bf Z_2}$ symmetry is not necessary from the view point of naturalness~\cite{Craig:2015pha,Barbieri:2016zxn}.
Here we simply set the twin QCD coupling to be $\widehat{g}_3 =1$.
The change of the value of $\widehat{g}_3$ allowed by naturalness leads to a 10\% effect for $\phi_B (T_C) /T_C$.
In addition, we take $X_{\widehat{t}}=0$ in our evaluation for the following reason. 
A non-zero $X_{\widehat{t}}$ tends to induce unwanted color-breaking vacua.  
In order to avoid the appearance of such vacua, 
larger soft masses are required, which reduces the ratio between the effective mass and the cubic term. 
Thus, with a non-zero $X_{\widehat{t}}$, $\phi_B (T_C) /T_C$ will get smaller compared to the case with a vanishing $X_{\widehat{t}}$.

Now the ratio $\phi_B (T_C) / T_C$ is determined by the twin stop soft parameters and the $U(4)$-breaking scale $f$.
Figure~\ref{fig:firstordersoftmass} shows the ratio $\phi_B (T_C) / T_C$ 
as the function of the left and right-handed twin stop (common)  soft masses $| \widetilde{m}^2_{\widehat{Q}} |= | \widetilde{m}^2_{\widehat{t}_R} |\equiv M_{\widehat{\rm stop}}^2$ for  $v_A/f=0.123$. The renormalization scale is set to be $\mu=T$.
We can see that the ratio $\phi_B (T_C) /T_C$ takes the maximal value for the 
massless limit of the 
light twin stop $M_{\widehat{\rm stop}}\simeq 0$, which is roughly 0.9. 
For other choices of the ratio $v_A/f \gtrsim 0.1$, required from the point of view of naturalness, 
we confirmed that 
$\phi_B (T_C) /T_C \simeq 0.9 > \widehat{g}_2$ for $M_{\widehat{\rm stop}}\simeq 0$.
Thus, for this parameter choice, the phase transition is the first order, which leads to the generation of the gravitational wave.
Note that here we admit the strong violation of the ${\bf Z}_2$ symmetry in the soft stop mass, 
but we assume that the ${\bf Z}_2$ symmetry is hold for $\tan \beta$ 
otherwise we cannot have the Mexican-hat type $U(4)$-breaking potential.

Now let us evaluate the spectrum of gravitational wave background generated 
in this model. 
For this purpose, we need to estimate the nucleation temperature $T_n$, 
the latent heat density $\alpha$ and the 
duration of the phase transition $\beta$ (see App.~\ref{sec:gravitational waves} for the detailed definition). 
They can be obtained by solving the bounce equations for the thermal resummed effective
potential $V(\phi_B,~T)= V_0 + V_{ CW} +V_{\rm thermal} + V_{\rm ring} + V_{\rm thermal}^{(2)}$.
Table~\ref{table:benchmark point}  shows the values of these parameters for our benchmark points, $\lambda =0.05,~\kappa_1=0.05,~\widehat{g}_3=1,~\tan \beta=10,~X_{\widehat{t}}=0$,
and $v_A/f=0.123$.

\begin{table}[tbp]
\centering
\begin{tabular}{|r|r|r|r|}
\hline
~~~~~~$T_n$ [GeV]  & ~~~$\phi_B (T_n)/T_n$ &$\alpha$~~~~  & ~~~~~~~$\beta / H(T_n)$~~\\
\hline 
$682$ & $1$~~~& ~~~$7 \times 10^{-3}$  & $7\times 10^4$~~\\
\hline
\end{tabular}
\caption{\label{table:benchmark point} Parameters $T_n,~\phi_B (T_n) / T_n,~\alpha$ and $\beta/ H(T_n)$ 
for the evaluation of the spectrum of gravitational wave background with the benchmark point  
$\lambda=0.05,~\widehat{g}_3=1,~\kappa_1 = 0.05,~M_{\widehat{\rm stop}}=0,~X_{\widehat{t}}=0,~v_A / f=0.123$ and $\tan \beta =10$.}
\end{table}
%
\begin{figure}[h]
\centering\includegraphics[width=9cm]{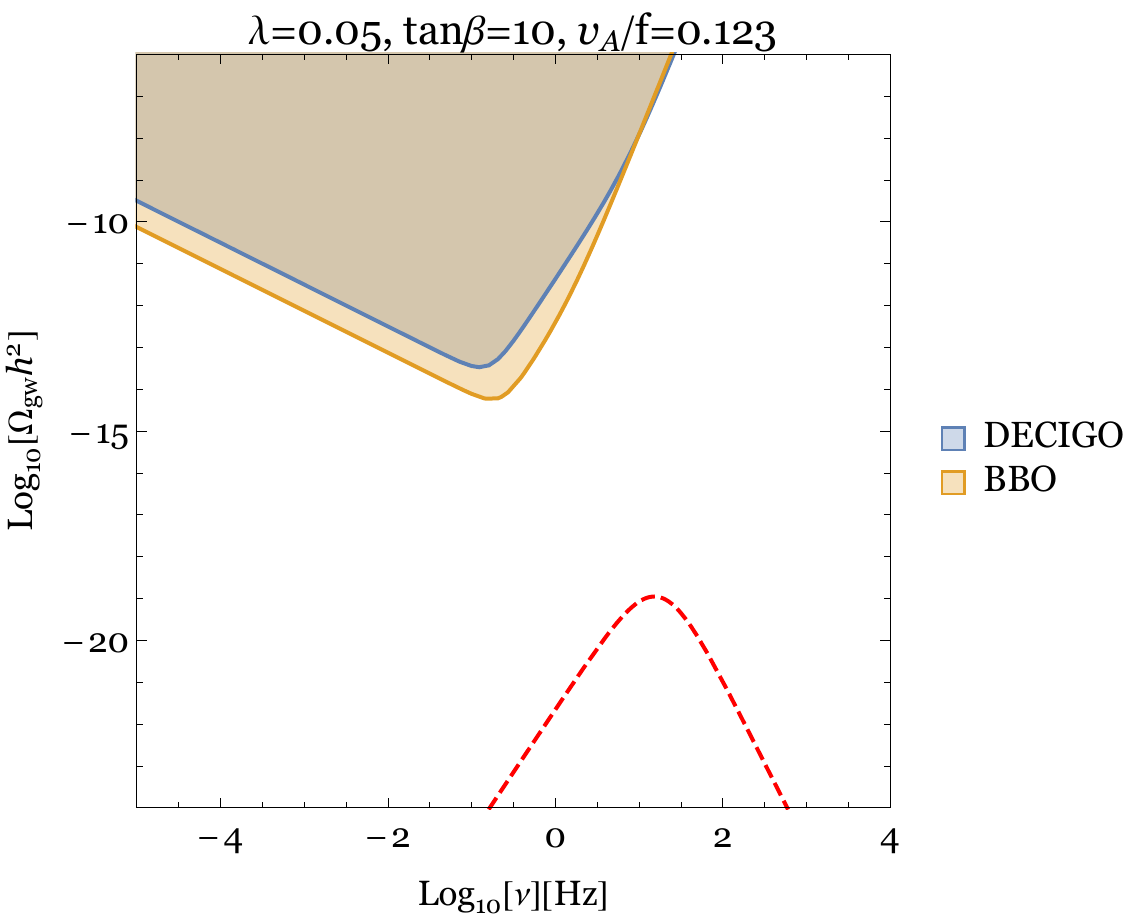}
\caption{The gravitational wave amplitude is shown. Orange and blue regions show the detectable regions by BBO and DECIGO, respectively.}\label{fig:figamplitude}
\end{figure} 
Figure~\ref{fig:figamplitude} shows the spectrum of the gravitational wave background for our 
benchmark point (see App.~\ref{sec:gravitational waves} for the formalism to calculate it).  
The most dominant source of the gravitational wave for our benchmark point is found to be the sound wave of the plasma bulk motion after the bubble collision, $\Omega_{\rm gw}h^2 \simeq \Omega_{\rm sw}h^2$ given by \eqref{eq:sound wave}.
The peak frequency is around $\mathcal{O} (10)$Hz and the peak amplitude of gravitational wave is around $\mathcal{O} (10^{-19})$ due to the large $\beta / H(T_n) \simeq 7\times 10^4$ and small $\alpha \simeq 7\times 10^{-3}$.
We can easily see that it is well below the sensitivities of DECIGO and BBO.
It was pointed out in Ref.~\cite{Ellis:2018mja} that the formula \eqref{eq:sound wave} overestimates the gravitational wave amplitude in the region of large $\beta/H$ (typically, $\beta / H \gtrsim 10^2$ for $\alpha > 5 \times 10^{-3}$). 
Thus it should be emphasized that the gravitational wave amplitude in Fig.~\ref{fig:figamplitude} is an upper bound 
and we expect that it will be much weaker in reality.

It is nontrivial whether our benchmark point, which gives the maximal ratio $\phi_B(T_C)/T_C$, 
gives the maximal amplitude of the gravitational wave background. 
We numerically confirmed that it is approximately maximal for our benchmark point. Concretely, 
\begin{itemize}
\item For $\lambda$, $\kappa_1$ and $M_{\widehat{\rm stop}}$, 
we confirmed that smaller $\lambda+\kappa_1$ and $M_{\widehat{\rm stop}}$ give
larger gravitational wave amplitude. Since we restrict them as $\lambda> 0.05,  \kappa_1>0.05$, 
and $M_{\widehat{\rm stop}}>0$,\footnote{Note that when we allow the negative twin stop soft masses, the gravitational wave amplitude would be larger. 
In this case, however, we have to take account of the $SU(3)_{\widehat{C}}$ breaking minimum hence we do not consider such a scenario in this paper.} 
our benchmark point gives the maximal amplitude. 

\item The peak amplitude of gravitational wave, $\Omega^{\rm peak}_{\rm gw}h^2$, does not depend on the breaking scale $f$.
We can write the effective potential as $V(\phi_B,~T,~f,~M_{\widehat{\rm stop}}) = T^4 \mu (\widetilde{\phi},~\widetilde{f},~\widetilde{M}_{\widehat{\rm stop}})$, where $\widetilde{\phi},~\widetilde{f}$ and $\widetilde{M}_{\widehat{\rm stop}}$ are parameters normalized by the temperature, $\widetilde{\phi}\equiv \phi /T,~\widetilde{f}\equiv f/T$ and $\widetilde{M}_{\widehat{\rm stop}} \equiv M_{\widehat{\rm stop}} /T$.
One can show that the bounce action $S_3 / T $ given by \eqref{eq:bounce action} is $S_3 / T = S_3 / T (\widetilde{\phi},~\widetilde{f},~\widetilde{M}_{\widehat{\rm stop}})$, after rescaling the radial coordinate as $r' = r / T$.
Then by definitions of $\alpha$ and $\beta$ parameters given by \eqref{eq:alpha parameter} and \eqref{eq:beta parameter}, we obtain $\alpha= \alpha  (\widetilde{\phi},~\widetilde{f},~\widetilde{M}_{\widehat{\rm stop}})$ and $\beta/H =\beta/H  (\widetilde{\phi},~\widetilde{f},~\widetilde{M}_{\widehat{\rm stop}})$.
The peak amplitude of gravitational wave, $\Omega_{\rm gw}^{\rm peak} h^2$, only depends on the $\alpha$ and $\beta / H$ parameters at $T= T_n$ hence we get $\Omega^{\rm peak}_{\rm gw} h^2 = \Omega^{\rm peak}_{\rm gw} h^2  (\widetilde{\phi},~\widetilde{f},~\widetilde{M}_{\widehat{\rm stop}})|_{T=T_n}$. 
The nucleation temperature is roughly given by $T_n \simeq T_B \simeq T_C$, where $T_B$ and $T_C$ are given in Sec.~\ref{sec:twin phase} and App.~\ref{sec:thermal effective potential}, respectively.
From the expression \eqref{eq:thermal massB}, we can easily find $f/T_n \simeq f/ T_B = {\rm const}$. 
In addition, from the expression \eqref{eq:strong first order}, the fraction $\phi_B (T_n) / T_n \simeq \phi_B (T_C) / T_C$ does not depend on the breaking scale $f$ (the quartic coupling $\xi$ is less sensitive to the change of $T_n$).
Thus, when we vary the breaking scale $f$ with $M_{\widehat{\rm stop}}=0$, the peak amplitude of gravitational wave does not change.
On the other hand, the peak frequency $\nu_{\rm peak}$ is proportional to the nucleation temperature, $T_n$, hence a smaller $f$ leads to a lower peak frequency due to the lower nucleation temperature.
We numerically confirm this behavior.

\item We numerically confirm that a smaller $\tan \beta$ makes the gravitational wave amplitude larger.
However, a smaller $\tan \beta$ leads to a larger up-type Higgs-top Yukawa coupling, $Y_{\widehat{t}} = y_{\widehat{t}} / \sin \beta$.
Here we impose the perturbative condition of the Yukawa coupling $Y_{\widehat{t}}^2 /(4 \pi)\lesssim1$ at the electroweak scale.
This condition gives $\tan \beta \gtrsim 0.28$.
For $\tan \beta =0.28$, the peak amplitude of gravitational wave is larger
than that of $\tan \beta=10$ by merely around factor 10.

\end{itemize}

 Note also that the change of the value of the twin QCD coupling constant allowed by naturalness affects the amplitude of gravitational wave by around factor 10 at most, and hence this effect does not change our result significantly.
Thus, we conclude that, even if we take the effect of a light twin stop into account, it is almost impossible to generate gravitational wave background detectable by DECIGO or BBO.

Finally, we would like to give some comments.
We have assumed that $\phi_B$ acquires the VEV first and $\phi_A$ does later. 
In order to verify this assumption, we have calculated the thermal resummed effective potential $V(\phi_A,~\phi_B,~T_n)$ for both of the Higgs fields $\phi_A$ and $\phi_B$ when $\phi_B$ gets the VEV at $T=T_n$.
We numerically confirmed that the potential minimum appears only in the $\phi_B$ direction at $T_n$ given in TABLE.~\ref{table:benchmark point}.
Therefore, the assumption of two-step phase transition is validated.

The resummed effective potential at finite temperature depends on a gauge-fixing parameter. In our calculation, we adopted the Landau gauge.
The effect of gauge dependence is discussed in, {\it e.g.}, Ref.~\cite{Wainwright:2011qy,Chiang:2017zbz}.
According to Ref.~\cite{Chiang:2017zbz}, the uncertainty due to gauge choice is roughly one or two order magnitude for $\Omega_{\rm gw}h^2$.
Even when we take this uncertainty into account, the gravitational wave amplitude shown in Fig.~\ref{fig:figamplitude} still does not reach the detectable regions by DECIGO and BBO. 
Therefore, our conclusion is still robust.

\section{Discussion and conclusion}\label{sec:discussion}

We have investigated the dynamics of the electroweak phase transition and the phase transition associated with global $U(4)$ breaking in twin Higgs models with and without supersymmetric completion.
In Sec.~\ref{sec:twin phase}, we found that the electroweak phase transition in twin Higgs models cannot be analyzed perturbatively as long as the effective potential is given by \eqref{eq:general} and \eqref{eq:finite EFT}.
It does not satisfy the condition of a strong first order phase transition, and hence we cannot expect for the realization of the electroweak baryogenesis 
as well as the generation of gravitational wave background.
In Sec.~\ref{sec:minimal}, we considered the $U(4)$-breaking phase transition in twin Higgs models without any UV completions such as composite Higgs and SUSY.
We confirmed that the $U(4)$-breaking phase transition is the first order only when $\lambda +\kappa_1 \lesssim 0.04$ is satisfied.
However, as discussed in Sec.~\ref{sec:SUSY twin Higgs}, we obtained the relation $\lambda + \kappa_1 >0.1$ in order to realize the adequate EWSB and the conditions $\lambda> \sigma_1 ,~\kappa_1,~\rho_1$.
Thus, the $U(4)$-breaking phase transition cannot be the first order, and  we expect that there is no gravitational wave production.
In Sec.~\ref{sec:transition with twin stop}, we considered the $U(4)$-breaking phase transition with supersymmetric UV completions in the decoupling limit where only the effect of light twin stops is taken into account.
We calculated the resummed effective potential including the dominant two-loop twin QCD contribution.
Then, we confirmed that the $U(4)$-breaking phase transition can be analyzed perturbatively only when the light twin stop masses with $M_{\widehat{\rm soft}} \simeq 0$ are realized.
We calculated the largest possible gravitational wave amplitude within the parameters for which the EWSB conditions and the conditions $\lambda >\sigma_1,~ \kappa_1,~\rho_1$ are satisfied.
However, we found that the gravitational wave amplitude cannot reach the detectable regions by DECIGO and BBO.

We conclude that it is impossible to produce large enough amplitude of gravitational wave to be detected by DECIGO or BBO in twin Higgs models, under our assumptions such as taking the decoupling limit, the perturbative conditions $\lambda>\sigma_1,~\kappa_1,~\rho_1$ and the trajectory of two-step phase transition.
We need to give a comment.
If there is an additional field strongly coupled to the Higgs fields $H_A$ and $H_B$, the dynamics of the electroweak phase transition and the $U(4)$-breaking phase transition will be changed due to the additional contribution to the effective potential.
For example, as mentioned in Sec.~\ref{sec:SUSY twin Higgs}, there is a singlet scalar field coupled to the Higgs field $H_A$ and $H_B$ in F-term twin Higgs models.
If such a singlet scalar field is sufficiently light during the $U(4)$-breaking phase transition, the situation might be dramatically changed.
In this paper, we do not consider such specific cases because we are mostly interested in giving model independent predictions.

\acknowledgments
We are grateful to Marcin Badziak, David Curtin, Masahiro Ibe, Ryusuke Jinno, Thomas Konstandin, Marek Lewicki, Satoshi Shirai, and Teruaki Suyama for helpful discussions and comments.
KF acknowledges the financial support from International Research
Center for Nanoscience and Quantum Physics, Tokyo Institute of
Technology.
KK is supported by IBS under the project code, IBS-R018-D1. 
YN thanks the Galileo Galilei Institute for Theoretical Physics
for the hospitality and the INFN for partial support during the completion of this work.
YN is supported by the DOE grant DE-SC0010008.
MY is supported in part by JSPS KEKENHI Grant Numbers JP25287054,
JP15H05888, JP18H04579, and JP18K18764.
KF and MY were supported by JSPS and NRF under the
Japan - Korea Basic Scientific Cooperation Program and would like to thank participants attending the JSPS and NRF conference for useful comments.
KK is grateful for the Mainz Institute for Theoretical Physics (MITP)
for the kind hospitality
and fruitful discussions,
where the present work is partially done during the MITP program
``Probing Baryogenesis via LHC and Gravitational Wave Signatures''  in
June 2018.

\appendix

\section{Finite temperature effective potential and phase transition}\label{app:finite temperature}

In this appendix, we give the details of the calculations used for evaluating
thermal potential 
and stochastic gravitational wave background from 
the first order phase transition in Secs.~\ref{sec:twin phase} and \ref{sec:U(4) breaking phase transition}. 
We also give the criteria to judge the validity of the perturbative calculation 
for thermal potential.

\subsection{Thermal effective potential and validity of perturbation theory}\label{sec:thermal effective potential}

Here we give the way to calculate the thermal potential used in 
Secs.~\ref{sec:twin phase} and \ref{sec:U(4) breaking phase transition}. 
We mainly follow the discussions in Ref.~\cite{Quiros:1999jp}.

The thermal effective potential is divided into three parts and can be schematically written as
\begin{align}
V_{\rm eff}&=V_0 + V_{\mathrm{CW}} + \Delta V_{\rm th}.\label{eq:ressumed effective potential}
\end{align}
Here $V_0$, $V_{\mathrm{CW}}$ and $\Delta V_{\rm th}$ represent the tree level potential, the one-loop Coleman-Weinberg potential 
and the thermal contributions respectively. 
Since we shall see that the perturbative calculation is not necessarily justified at high-temperature, we will later take into account higher order effects partially to improve the perturbativity.

The Coleman-Weinberg potential $V_{\rm CW}$ is given by 
\begin{align}
V_{\mathrm{CW}} = \sum_i (-)^{F_i}\frac{n_i}{64\pi^2} m_i^4 (\phi) \left(\log \left(\frac{m_i^2 (\phi)}{\mu^2}\right) -\frac{3}{2}\right), 
\end{align}
where $n_i$ is the number of degrees of freedom of a particle $i$, $m_i(\phi)$ represents the field dependent mass of  the particle $i$, and $\mu$ is a renormalization scale, and $(-)^{F_i}$ gives 1 for bosons and $-1$ for fermions.
Here we adopt the $\overline{\rm DR}$ renormalization scheme.

Thermal contributions to the effective potential include the one-loop effective potential given by
 \begin{align}
 V_{\mathrm{thermal}} &= \sum_{i} \left(\frac{n_{Bi}T^4}{2\pi^2 }J_B [m_{Bi}^2 (\phi) /T^2]+\frac{n_{Fi} T^4}{2\pi^2} J_F [m_{Fi}^2 (\phi) /T^2] \right), \label{tCW} \\
 J_B [m^2 (\phi)/T^2]&=\int^{\infty}_{0}dx x^2 \log\left(1-e^{-\sqrt{x^2+m^2 (\phi) /T^2}}\right), \label{JB}\\ 
 J_F [m^2 (\phi)/T^2]&=\int^{\infty}_{0}dx x^2 \log\left(1+e^{-\sqrt{x^2+m^2 (\phi)/T^2}}\right), \label{JF}
\end{align}
where $i$ runs the particle species and the suffixes $B$ and $F$ represent Boson and Fermion contributions, respectively. 
We here adopt the imaginary time formalism. For later use, we note that at high temperature $m^2 (\phi)/T^2 \ll 1$, they are approximated as
\begin{align}
 J_B [m^2 (\phi) /T^2] &= -\frac{\pi^4}{45}+\frac{\pi^2}{12}\frac{m^2 (\phi)}{T^2}-\frac{\pi}{6}\left(\frac{m^2 (\phi)}{T^2}\right)^{\frac{3}{2}} - \frac{m^4 (\phi)}{32 T^4}\log \left(\frac{m^2 (\phi)}{a_b T^2}\right), \label{eq:boson}\\
 J_F [m^2 (\phi)/T^2] &= \frac{7\pi^4}{360}-\frac{\pi^2}{24}\frac{m^2 (\phi)}{T^2} - \frac{m^4 (\phi)}{32 T^4}\log\left(\frac{m^2 (\phi)}{a_f T^2}\right), \label{eq:fermion}
\end{align}
with
\begin{equation}
 a_b = 16\pi^2 \exp\left[\frac{3}{2}-2\gamma_E\right], \quad
 a_f = \pi^2 \exp\left[\frac{3}{2}-2\gamma_E\right],
\end{equation}
where $\gamma_E$ is the Euler's constant. 

However, it will be immediately seen that this perturbative expansion breaks down at high 
temperature. 
The quadratic divergent contributions to the self-energy from the $n$-loop diagram, often called as   the ring diagram or daisy diagram~\cite{Arnold:1992rz},  behave as~\cite{Fendley:1987ef}
\begin{equation}
a^2 \frac{T^3}{m(\phi)} \left(\frac{a T^2}{m^2(\phi)}\right)^{n-1}, 
\end{equation}
where $a$ is a constant determined by the coupling constants, 
which are the expanding parameters in the zero-temperature perturbative calculations. 
Thus, for $a T^2/m^2(\phi) \gg 1$, the perturbative calculation is not valid especially in 
calculating the critical temperature at the phase transition. 

By taking a closer look at the structure of the divergences, we can see that 
they come from the Matsubara zero mode of bosonic particles. 
Thus this problem is relaxed by ``resumming'' the ring diagrams of bosonic particles 
where we replace the mass of the bosonic particle $m_{Bi}(\phi)$ in the one-loop thermal potential by the dressed one,
\begin{equation}
\overline{m}_{Bi}^2 (\phi,~T) \sim m_{Bi}^2 (\phi) +\Pi_i(T), \quad \Pi_i(T)=cT^2, \label{eq:thermal field dependent mass}
\end{equation}
where $\Pi_i(T)$  is the one-loop self energy of the bosonic particle 
corresponding to the ring diagrams. 
Here  $c$ denotes the contribution of gauge and Yukawa couplings. 
This is equivalent to adding 
\begin{align}
V_{\mathrm{ring}}=-\frac{n_{Bi}T}{12\pi}\left(\left(\overline{m}^2 (\phi,~T)\right)^{\frac{3}{2}}-\left(m^2 (\phi)\right)^\frac{3}{2} \right), \label{eq:ressumation}
\end{align}
to the thermal potential so that
\begin{equation}
\Delta V_{\rm th} = V_{\rm thermal}+V_{\rm ring}. 
\end{equation}
After the resummation, the $n$-loop quadratically divergent diagram behaves as 
\begin{equation}
a^{n+1} \frac{T^{2n+1}}{m^{2n-1}(\phi)}\left(\frac{a T}{m (\phi)}\right). 
\end{equation}
Thus for $a<1$, the condition for the perturbative expansions to be validated is improved as
\begin{equation}
\frac{aT}{m(\phi)}\ll 1. 
\end{equation}

When non-Abelian gauge fields are involved, we need to take into account another subtle 
issue. Although the transverse modes of the gauge fields are massless at the one-loop 
perturbative calculation, 
it is known that through the non-perturbative process it receives the so-called
magnetic mass, $\sim g^4 T^2$, with $g$ being the gauge coupling. 
Then, with a similar discussion given above, the higher loop of non-Abelian gauge bosons 
will give the contributions with the powers of $g^2 T/m(\phi)$~\cite{Linde:1980ts,Gross:1980br}
and the perturbation breaks down at high temperature~\cite{Arnold:1994bp}, 
\begin{align}
g^2 \frac{T}{m(\phi)} >1. 
\end{align}
In this case, even the resummed effective potential \eqref{eq:ressumed effective potential}
is not reliable and the dynamics of phase transition should be analyzed by lattice simulations. 
Since we expect the parameter $a$ given above is at most unity, 
we conclude that the resummed effective potential is valid for $\gamma_2 \equiv g^2 T/m(\phi) \simeq  g T/ \phi <1$ when $m(\phi) \simeq g \phi$.

Let us now give our criteria for a first order phase transition and a "strong" first order phase transition to occur. 
Starting from the one-loop effective potential \eqref{eq:ressumed effective potential}, 
we can approximate it as 
\begin{align}
V = \frac{1}{2}M^2(T)\phi^2 -ET\phi^3+\frac{\xi(T)}{4}\phi^4. \label{eq:approximate potential}
\end{align}
Here $M^2 (T),~E$ and $\xi(T)$ represent a temperature 
dependent mass, a numerical coefficient depending on coupling constants, and a temperature dependent self-coupling, respectively. 
Note that the coefficient $E$ comes from the loops from bosonic particles (see Eqs.~\eqref{eq:boson} and~\eqref{eq:fermion}). 
The thermal potential \eqref{eq:approximate potential} can have two minima. 
One is at the origin, while the other is not, 
depending on the temperature and other model parameters. 
As the temperature decreases, the two minima can get 
degenerated.
We define the temperature at which the two minima degenerate 
as the critical temperature, $T_C$. 
At that temperature, the effective potential \eqref{eq:approximate potential} 
is written as
\begin{align}
V_{T= T_C} = \frac{\xi(T_C)}{4} \phi^2 (\phi - \phi(T_C))^2 ,
\end{align}
where $\phi(T_C) \neq 0$ is the other minimum at $T_C$.
Or we can write 
\begin{align}
\frac{T_C}{\phi(T_C)} = \frac{\xi (T_C)}{2 E}. \label{eq:strong first order}
\end{align}
Below the critical temperature, the minimum other than the origin is energetically favored 
and hence tunneling from the origin (symmetric phase) to the other minimum (broken phase)
can occur and the symmetry breaks down. 
We have seen that at the small field values, including the origin, 
the resummed effective potential~\eqref{eq:ressumed effective potential} or 
its approximated one~\eqref{eq:approximate potential}  is not reliable. 
However, we here give the criteria that the perturbative calculation is allowed to use 
and the symmetry breaking is first order 
if the potential minimum in the broken phase 
satisfies the condition that the perturbative calculation is valid, $gT_C /\phi (T_C) <1$,
since the tunneling rate is determined mainly by the information of the potential 
around the broken phase but not the symmetric phase. 
Moreover, we define the phase transition as strong first order if 
\begin{equation}
\frac{T_C}{\phi(T_C)}<1,\label{eq:sphaleron decoupling}
\end{equation}
is satisfied.
The difference between first order phase transition ($gT_C /\phi (T_C) <1$) and strong first order phase transition ($T_C / \phi (T_C)<1$) is important when we consider electroweak baryogenesis because the sphaleron decoupling condition is given by Eq.~\eqref{eq:sphaleron decoupling}.
On the other hand, this difference is not important when we discuss the gravitational wave background generated by first order phase transition.
Since a first order phase transition proceeds through bubble nucleation, the production of gravitational wave background requires a first order phase transition, not a strong first order phase transition as we will see later.
From \eqref{eq:strong first order}, we can see that a strong first order phase transition
takes place if the self-coupling $\xi (T_C)$  is small enough and 
the cubic prefactor $E$ is large enough. 
This is because the parameter $E$ determines the height of the barrier between 
the origin and the other minimum. 
Since the cubic term comes from the bosonic loop contribution, 
bosons strongly coupled to $\phi$ are needed for a strong first order phase transition. 

Before closing this subsection, let us comment on the effect of the ring diagram on the strength
of the phase transition. 
In the expression of the ring diagram contribution \eqref{eq:ressumation},
the thermal field dependent mass $\overline{m}(\phi ,~T)$ 
is roughly given by Eq.~\eqref{eq:thermal field dependent mass}.
After the resummation,
if the thermal mass is much larger than the zero-temperature part, that is, $m^2_B(\phi ) \ll \Pi$ at $T=T_C$, 
$(\overline{m}^2)^{\frac{3}{2}}$ behaves like a constant term $\simeq \Pi^{\frac{3}{2}}(T_C) $ which does not give the potential barrier.
This effect makes $\phi(T_C) $ small hence the resummation generally makes the phase transition weaker.

\subsection{Phase Transition and Gravitational Waves} \label{sec:gravitational waves}

In this appendix, we review how a first order phase transition proceeds and 
the stochastic gravitational wave background generated from it is evaluated. 

A first order phase transition occurs as a result of true vacuum bubble nucleations. 
This is understood as quantum or thermal tunneling from a false vacuum to a true vacuum that is separated by a potential barrier.
The tunneling rate or the bubble nucleation rate $\Gamma(T)$ per unit volume and unit time at finite temperature is evaluated as~\cite{Linde:1980tt} 
\begin{align}
\Gamma (T) &=A(T) e^{-S_3 /T},  \label{eq:tunneling rate}\\ 
\frac{S_3}{T}&= \int_{0}^{\infty} dr 4 \pi r^2 \left( \frac{1}{2}\left(\frac{d \phi(r)}{dr} \right)^2 +V(\phi(r),~T) \right), \label{eq:bounce action}
\end{align}
where the prefactor $A(T) \sim T^4$ is determined by the quantum effects, $S_3$ represents the $O(3)$ symmetric bounce action and 
$\phi(r)$ is the  ``bounce solution'' of the following equation of motion 
\begin{align}
\frac{d^2 \phi}{d r^2}+\frac{2}{r}\frac{d \phi}{dr} - \frac{\partial V(\phi,~T)}{\partial \phi}=0 ,\label{eq:EOM}
\end{align}
with boundary conditions
\begin{align}
\phi(r\to \infty)=\phi_{\mathrm{False}},~ \left. \frac{d \phi}{dr} \right|_{r=0} =0 \label{eq:bounce boundary}.
\end{align}
Here $r$ is the radial coordinate in the three dimensional polar coordinate system and 
$\phi_{\mathrm{False}}$ is the field value of the false vacuum.

The time or the temperature of the phase transition is characterized
by the nucleation time $t_n$ or the temperature $T_n$,  
defined as a temperature when the nucleation 
probability inside one Hubble volume $H^3 (T)$  gets unity, 
\begin{align}
\int^{t_{n}}_{0} \frac{\Gamma (T)}{H^3 (T)}dt=\int^{\infty}_{T_n} \frac{dT}{T} \frac{\Gamma(T)}{H^4 (T)}=1. \label{eq:nucleation temperature}
\end{align}
Since the dominant contribution in the integral \eqref{eq:nucleation temperature} 
comes from that around $t \sim t_n$ or $T\sim T_n$, it can be approximated as
\begin{align}
\frac{\Gamma (T_n)}{ H^4(T_n)} =1,  \label{eq:app nucleation temperature}
\end{align}
which can be used to determine $T_n$.
Since the EWSB takes place at $T_n\sim {\cal O}(100)$ GeV, 
the bounce action at the time of bubble nucleation is roughly given by 
\begin{align}
\frac{S_3}{T_n} = 4\log\left(\frac{T_n}{H}\right)\sim 140.
\end{align}
Generally speaking, the nucleation temperature is lower than the critical temperature $T_n < T_C$.
In order to determine the nucleation temperature as well as the bubble profile accurately, 
we need to solve the equation of motion \eqref{eq:EOM} with the boundary condition
\eqref{eq:bounce boundary} numerically. 
We here adopt a method dubbed as the under/over-shooting method, 
developed in Ref.~\cite{Cai:2017tmh}.

Now let us give the expressions of the spectrum of the gravitational background 
from the first order phase transition. 
Since the broken phase is energetically favored, the nucleated bubbles expand, 
and collide each other, and finally the whole Universe settles down to the true vacuum. 
Since the bubble collisions as well as the plasma bulk motion induced by the bubble dynamics are highly inhomogeneous and violent process, gravitational waves are emitted through such processes.

The spectrum of the gravitational wave is determined by the (initial) kinetic energies of the bubbles and the duration of the phase transition. 
The former is provided by the latent heat 
density~$ \Delta \rho = \rho(\phi_{\mathrm{False}},T_n) -\rho(\phi_{\mathrm{True}},T_n)$, 
where $\rho(\phi, T)$ is the thermodynamic internal energy, but not the potential energy. 
By identifying the effective thermal potential with the free energy, 
$\mathcal{F}(\phi,T)= V(\phi,T)$, we obtain
\begin{equation}
\rho(\phi,T)=\mathcal{F}(\phi,T) +sT = \mathcal{F}(\phi,T) -T \frac{d}{dT} \mathcal{F}(\phi,T) 
\end{equation}
so that 
\begin{align}
\Delta \rho = \Delta V(T)-T\frac{d}{dT} \Delta V(T),\quad \Delta V= V(\phi_{\mathrm{False}},T)-V(\phi_{\mathrm{True}},T).
\end{align}
We parameterize the kinetic energy of bubbles by a dimensionless parameter $\alpha$ representing the ratio between the latent heat density and the radiation energy density,
\begin{align}
\alpha = \frac{\Delta \rho}{\rho_{\mathrm{rad}}},  \label{eq:alpha parameter}
\end{align}
where $\rho_{\rm rad} = g_*  \pi^2 T_*^4/30$ denotes the energy density of radiation. 
Here $T_* \simeq T_n$ is the temperature at which the gravitational waves are emitted. 
The duration of the phase transition is characterized by the parameter $\beta$, defined by
 \begin{align}
 \Gamma(t) \simeq  \Gamma_0 e^{\beta t}, \label{eq:beta parameter}
 \end{align}
with $\Gamma_0$ being a constant. $\beta$ is expressed in terms of the bounce action as
\begin{align}
\frac{\beta}{H} = T\frac{d}{dT}\left(\frac{S_3}{T} \right).
\end{align}

It has been argued that not only the bubble collision or the scalar field dynamics, 
but also the plasma dynamics caused by the bubble dynamics source the gravitational waves~\cite{Turner:1990rc,Kosowsky:1991ua,Kosowsky:1992rz,Kosowsky:1992vn,Turner:1992tz}. It is indeed
found to be the dominant contribution to the gravitational wave background 
since due to the interaction between the scalar field bubble wall and the plasma,
the energy originally carried by bubble walls is quickly taken away to the plasma bulk motion. 
According to the popular convention, we further classify it into the sound waves in the plasma described in the linear regime, 
which are generated by the bubble motion and 
generate gravitational waves around the bubble collision, and the turbulence of plasma bulk motion further developed in the non-linear regime after the bubble collision.
Then the total contribution can be schematically written as
\begin{align}
\Omega_{\mathrm{gw}} h^2=  \Omega_{\mathrm{bubble}} h^2 + \Omega_{\mathrm{sw}} h^2 +\Omega_{\mathrm{tur}} h^2 \simeq \Omega_{\mathrm{sw}} h^2 +\Omega_{\mathrm{tur}} h^2, 
\end{align}
where $\Omega_{\mathrm{bubble}},~\Omega_{\mathrm{sw}}$ and $\Omega_{\mathrm{tur}}$ denote the contributions from the bubble collisions, sound waves and turbulence of the plasma, respectively.

For the contributions from sound waves, we adopt the expressions in Ref.~\cite{Caprini:2009yp,Binetruy:2012ze,Hindmarsh:2015qta,Caprini:2015zlo} as,
\begin{align}
\Omega_{\mathrm{sw}} h^2 (\nu)&= 2.65 \times 10^{-6} \left(\frac{H(T_n)}{\beta}\right) \left(\frac{\kappa_v \alpha}{1+\alpha}\right)^2 \left(\frac{100}{g_*}\right)^{\frac{1}{3}} v_b  \left(\frac{\nu}{\nu_{\mathrm{sw}}}\right)^3 \left(\frac{7}{4+3\left(\frac{\nu}{\nu_{\mathrm{sw}}}\right)^2}\right)^{\frac{7}{2}}, \label{eq:sound wave} \\
\kappa_v &= \frac{\alpha}{0.73+0.083 \sqrt{\alpha} +\alpha} ,\\
\nu_{\mathrm{sw}}&= 1.9 \times 10^{-5} \mathrm{Hz} \frac{1}{v_b} \left(\frac{\beta}{H(T_n)}\right) \left(\frac{T_n}{100 \mathrm{GeV}}\right) \left(\frac{g_*}{100}\right)^{\frac{1}{6}},
\end{align}
with $\kappa_v $ 
being the fraction of vacuum energy that gets converted into the fluid kinetic energy.
$v_b$ is the bubble wall velocity.
For the contributions from the turbulence plasma, 
\begin{align}
\Omega_{\mathrm{tur}}  h^2 (\nu)&= 3.35\times 10^{-4} \left(\frac{H(T_n)}{\beta}\right) \left(\frac{\kappa_{\rm tur} \alpha}{1+\alpha}\right)^{\frac{3}{2}} v_b \frac{\left(\frac{\nu}{\nu_{\mathrm{tur}}} \right)^3}{\left(1+\frac{\nu}{\nu_{\mathrm{tur}}}\right)^{\frac{11}{3}} \left(1+\frac{8\pi \nu}{H_0} \right)},\\
\kappa_{\rm tur}&\simeq 0.1 \times \kappa_{v},\\
\nu_{\mathrm{tur}}&=2.7 \times 10^{-5} \mathrm{Hz} \frac{1}{v_b} \left(\frac{g_*}{100}\right)^{\frac{1}{6}} \left(\frac{T_n}{100 \mathrm{GeV}}\right) \left( \frac{\beta}{H(T_n)}\right).
\end{align}
The estimate for the bubble wall velocity has ambiguities, but here 
we assume the so-called detonation and adopt the formula in Ref.~\cite{Steinhardt:1981ct}, 
\begin{align}
v_b = \frac{1/\sqrt{3} +\sqrt{\alpha^2 +2\alpha/3 }}{1+\alpha}, 
\end{align}
so that it gives the maximal estimate for the amplitude of the gravitational wave background.
In our setup the latent heat density is small, $\alpha ={\cal O}(10^{-3 \sim -2})$. 
If the bubble wall velocity is smaller and in the deflagration regime, the 
gravitational wave background is much smaller.
It should be also noted that the formula of the gravitational wave coming from the sound waves, \eqref{eq:sound wave}, does not necessarily work and is likely to overestimate the gravitational wave amplitude for large $\beta / H$ (typically, $\beta /H >10^2$ for $\alpha >5 \times10^{-3}$)~\cite{Ellis:2018mja}.
Thus, to be precise, our estimate based on the formula \eqref{eq:sound wave} should be regarded as the upper bound of the gravitational wave amplitude.

Note that  by adopting the formula with the envelope approximation in Ref.~\cite{Huber:2008hg} 
the contributions from bubble collisions turned out to be subdominant in our setup.\footnote{
Recently it is claimed that there might be another contribution from ``fluid bubble'' in Refs.~\cite{Jinno:2016vai,Jinno:2017fby,Jinno:2017ixd}, but it gives a subdominant contribution compared to the sound waves~\cite{Hindmarsh:2013xza,Giblin:2014qia,Hindmarsh:2015qta}.} Thus we safely omitted the contributions from
the bubble collisions in our plots.

\bibliographystyle{JHEP}
\bibliography{twinref}

\end{document}